\documentclass[useAMS,usenatbib]{mn2e}

\usepackage{graphicx}
\usepackage{float}
\usepackage{placeins} 
\usepackage{paralist}
\usepackage{subfigure}
\usepackage{amssymb}

\newcommand{\HI}{H\,{\sc i}}

\newcommand{\Msun}{$M_{\odot}$}
\newcommand{\kms}{km\,s$^{-1}$}

\title{\HI-deficient galaxies in intermediate density environments}
\author[H. D\'{e}nes, V. A. Kilborn, B. S. Koribalski, O. I. Wong]
  {H.~D\'{e}nes$^{1, 2}$\thanks{E-mail: hdenes@astro.swin.edu.au}, V. A. Kilborn$^1$, B. S. Koribalski$^2$, O. I. Wong$^{3}$
   \\
  $^1$Centre for Astrophysics \& Supercomputing, Swinburne University of Technology,  PO Box 218, Hawthorn, VIC 3122, Australia\\
  $^2$Australia Telescope National Facility, CSIRO Astronomy and Space Science, P.O. Box 76, Epping, NSW 1710, Australia\\
  $^3$International Centre for Radio Astronomy Research, The University of Western Australia M468, 35 Stirling Highway, Crawley,\\  \hspace{2mm} WA 6009, Australia\\
  }
\date{Released 2002 Xxxxx XX}

\pagerange{\pageref{firstpage}--\pageref{lastpage}} \pubyear{2002}

\def\LaTeX{L\kern-.36em\raise.3ex\hbox{a}\kern-.15em
    T\kern-.1667em\lower.7ex\hbox{E}\kern-.125emX}

\begin{document}

\label{firstpage}

\maketitle

\begin{abstract}

Observations show that spiral galaxies in galaxy clusters tend to have on average less neutral hydrogen (\HI) than galaxies of the same type and size in the field. There is accumulating evidence that such \HI-deficient galaxies are also relatively frequent in galaxy groups. An important question is, which mechanisms are responsible for the gas deficiency in galaxy groups. To gain a better understanding of how environment affects the gas content of galaxies, we identified a sample of six \HI-deficient galaxies from the \HI\ Parkes All Sky Survey (HIPASS) using \HI-optical scaling relations. One of the galaxies is located in the outskirts of the Fornax cluster, four are in loose galaxy groups and one is in a galaxy triplet. We present new high resolution \HI\ observations with the Australia Telescope Compact Array (ATCA) of these galaxies. We discuss the possible cause of \HI-deficiency in these galaxies based on \HI\ observations and various multi-wavelength data. We find that the galaxies have truncated \HI\ disks, lopsided gas distribution and some show asymmetries in their stellar disks. We conclude that both ram pressure stripping and tidal interactions are important gas removal mechanisms in low density environments. 

\end{abstract}

\begin{keywords}
galaxies: evolution -- galaxies: general -- radio lines: galaxies -- surveys.
\end{keywords}

\section{Introduction}

One of the key questions in astrophysics is to understand the role of environment in galaxy evolution. We have evidence that the environment of a galaxy can play a major role in its evolution. Examples for this include the `morphology-density relation' \citep{Dressler1980, Fasano2000, Goto2003}, which refers to the increasing fraction of early-type galaxies in dense environments and the `star formation-density relation' \citep{Lewis2002, Kauffmann2004},which shows the suppression of star formation in high density environments. In terms of the gas content, spiral galaxies in high density environments (galaxy clusters) tend to have on average, less neutral hydrogen (\HI) than galaxies of the same type and size in the field (e.g. \citealt{Davis1973, Giovanelli1985, Solanes2001}). These observations imply that the gas reservoir of late-type galaxies in high density regions is removed, or consumed and  they are transformed into gas-poor quiescent galaxies (e.g. \citealt{Gunn1972, Fasano2000, Bekki2002, Bekki2011}). \HI\ observations can be used to investigate the environmental effects that may be responsible for transforming galaxies, since \HI\ is a good tracer of galaxy interactions. The \HI\ disk of a galaxy is relatively diffuse and usually a few times more extended than the stellar disk, which means it can be very easily disturbed by external forces \citep{Hibbard1996}. External processes can affect the overall \HI\ content of a galaxy (e.g. by gas removal or gas accretion) and they can affect the \HI\ morphology. In this paper we focus on gas removal processes. Typical observational signatures of gas removal are \HI\ tails and bridges, displaced \HI\ from the disk, lopsided and truncated \HI\ disks (e.g. \citealt{Chung2007, Hibbard2001}).

There are two main types of gas removal mechanisms: 
\begin{inparaenum} 
\item gravitational interactions between galaxies (tidal interactions, harassment, mergers); and 
\item hydrodynamical interactions between galaxies and the intra cluster medium (ICM) (ram pressure stripping).
\end{inparaenum} 
The main difference in the observational signatures of gravitational and hydrodynamical interactions is that gravitational interactions usually disturb both the gas and the stellar component of a galaxy, whereas interactions with the ICM mainly disturb the gas content. Thus, one method to distinguish between different gas removal mechanisms is to compare the optical morphology with the \HI\ morphology and surface brightness. The following signatures are expected for the different gas removal processes:
\begin{compactitem}
	\item Tidal interactions can cause arms, streams, warps and an asymmetric distribution in the \HI\ content (e.g. \citealt{Yun1994, Hibbard1996, Putman1998}). They can also cause stellar arms, streams, bars and warps and trigger enhanced star formation (e.g. \citealt{Toomre1972, Barnes1992, Larson1978, Kennicutt1987}). 
	\item Ram pressure stripping can produce truncated \HI\ disks, one sided \HI-tails, lopsided \HI\ morphology and enhanced star formation on the leading side of the galaxy (e.g. \citealt{Cayatte1994, Vollmer2001}). 
	\item Starvation occurs when a galaxy enters a dense environment such as a galaxy group or a galaxy cluster and the hot gas in the galaxy's halo is removed by ram pressure stripping or tidal interactions. This is expected to produce a normal sized, symmetric, low surface brightness \HI\ disk with no significant effect on the stellar disk (e.g. \citealt{Cayatte1994, Koopmann2004}).
\end{compactitem}

The effect of environment on the \HI\ content of galaxies is well studied in high density environments, especially in the Virgo cluster (e.g. \citealt{Kenney2004, Vollmer2001, Chung2009}). Several \HI-deficient galaxies - galaxies that have significantly less \HI\ than same type and sized galaxies in the field \citep{Haynes1983} - have also been identified in other galaxy clusters such as the Fornax \citep{Waugh2002, Schroder2001}, Coma \citep{Bravo-Alfaro2000}, Hydra \citep{Solanes2001}, and Pegasus \citep{Solanes2001, Levy2007} clusters. \HI-deficient galaxies were also found in compact groups \citep{Huchtmeier1997, Verdes-Montenegro2001}, where the main environmental effects seem to be tidal interactions \citep{Rasmussen2008} due to the high galaxy density. However environmental effects are not limited to high density environments. Pre-processing in galaxy groups has increasingly showed to play an important role in galaxy evolution. In recent years several studies identified \HI-deficient galaxies in loose galaxy groups \citep{Chamaraux2004, Kilborn2005, Sengupta2006, Kilborn2009, Westmeier2011, Hess2013}. Statistical studies of large galaxy samples also show that galaxies in intermediate density environments have less \HI\ compared to galaxies in the field (e.g. \citealt{Fabello2012, Hess2013, Catinella2013}). 
 
It is still a question, which gas removal mechanisms are dominant in galaxy groups. Since the physical properties of galaxy groups are different compared to galaxy clusters - e.g. lower velocity distribution, lower density of members, lower temperature of the intra group matter (IGM) - tidal interactions (e.g. \citealt{Kern2008, Rasmussen2008}) or a combination of ram pressure stripping and tidal interactions (e.g. \citealt{Davies1997, Mayer2006, Rasmussen2012}) are generally considered the most likely cause of \HI-deficiency. However some galaxy groups have a hot X-ray emitting IGM, which could play an important role in stripping off the cold gas. \cite{Sengupta2006} and \cite{Kilborn2009} show that groups with hot IGM have more \HI-deficient galaxies than groups without hot gas. In these groups ram pressure stripping could also play a significant role in influencing the gas content of galaxies. Further support to this is shown by \cite{Westmeier2011, Westmeier2013}, who found signs of ram pressure stripping in two galaxies of the Sculptor group. 

To better understand how the environment affects the gas content of galaxies, it is important to investigate the cause of gas-deficiency in individual galaxies. For this we have identified a sample of six strongly \HI-deficient galaxies from the \HI\ Parkes All Sky Survey (HIPASS, \citealt{Barnes2001}) using \HI-optical scaling relations from \cite{paper1} (see section~\ref{Sample_selection} and Fig.~\ref{fig:scaling-relation}). This sample differs from previous studies of \HI-deficient galaxies, as none of the sample galaxies are in dense cluster cores. The sample includes a galaxy in the outskirts of the Fornax cluster, four galaxies in various loose groups and one galaxy in a triplet. We present high-resolution ATCA observations of these \HI-deficient galaxies. In section~\ref{Data} we describe the sample selection, the \HI\ observations and the multi-wavelength archival data used in this work. In section~\ref{Results} we discuss the properties of our galaxies and the most likely causes of their \HI-deficiency. In section~\ref{Discussion} we compare our sample to known \HI-deficient galaxies in the literature and in section~\ref{Summary} we summarise our results.

Throughout this paper we use $H_{0}=70$ \kms\ Mpc$^{-1}$.  

\section{Data}
\label{Data}

\subsection{Sample selection}
\label{Sample_selection}

\begin{figure}
\includegraphics[width=84mm]{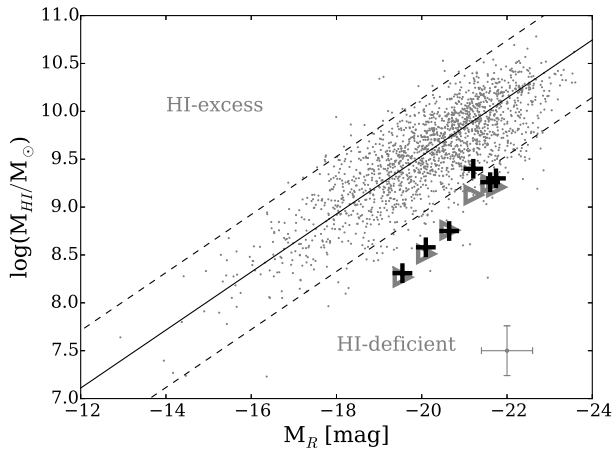}
\caption{\textit{R}-band magnitude versus \HI\ mass for the parent sample. The solid line shows the \textit{R}-band scaling relation from \citep{paper1} and the dashed line marks \textit{DEF} $= \pm 0.6$. Grey triangles show the original HICAT \HI\ data for the 6 sample galaxies and black crosses show the re-measured \HI\ mass from the HIPASS data cubes (see Table~\ref{tab:HI-properties}). The error bar displayed in the bottom right indicates the average uncertainties in the data.}
\label{fig:scaling-relation}
\end{figure}

To select a sample of \HI-deficient galaxies, the \textit{R}-band magnitude scaling relation from \cite{paper1} is applied to galaxies in the HIPASS Optical counterparts catalogue (HOPCAT, \citealt{Doyle2005}) to calculate their expected \HI\ mass. HOPCAT contains optical identifications for about 4000 HIPASS sources, however because of the 15'.5 spatial resolution of HIPASS some \HI\ detections correspond to more than one galaxy or to a whole galaxy group. To avoid confusion, we only consider \HI\ sources for the sample selection which can clearly be associated with a single galaxy. This gives a parent sample of 1798 galaxies for the selection. Fig.~\ref{fig:scaling-relation} shows the \HI-optical correlation for these galaxies and the \textit{R}-band scaling relation from \cite{paper1}. We calculate the expected \HI\ mass and the \HI-deficiency parameter for all galaxies, to measure their relative \HI\ content. The \HI-deficiency parameter ($DEF$, \citealt{Haynes1983}) is defined as: 
\begin{equation}
 \textsc{Def}_{HI} = \rmn{log[M}_{HI exp}] - \rmn{log[M}_{HI obs}], 
\end{equation}
 where M$_{HI exp}$ is the expected \HI\ mass, usually calculated from \HI\ scaling relations, and M$_{HI obs}$ is the measured \HI\ mass. Positive values show galaxies that have less \HI\ than the average and negative values show galaxies that have more \HI\ than the average. In the literature galaxies are usually considered to be \HI-deficent if $DEF > 0.3$, which corresponds to 2 times less \HI\ than the average galaxy in a low density environment. However, in order to study the most \HI-deficient galaxies and considering the $\sim$0.3 dex scatter of the scaling relations, we adopt a more conservative $DEF > 0.6$ threshold for \HI-deficiency. We find that 94 (5 \%) galaxies in HOPCAT are \HI-deficient above this threshold and therefore have at least 4 times less \HI\ than expected. From these we select the 6 most \HI-deficient late-type galaxies with $\delta < -30^{\circ}$ for high resolution \HI\ observations (grey triangles in Fig.~\ref{fig:scaling-relation}). These galaxies have \HI\ masses between 2 to 25.3$\times 10^{8}$\Msun\ and roughly 3 to 12 times less \HI\ than expected from the \textit{R}-band scaling relation. Table~\ref{tab:properties_summary} presents an overview of the basic properties of these galaxies.

\begin{table*}
\caption{Properties of the target galaxies. Stellar masses are calculated following \citep{Bell2001} and \HI\ masses are calculated from the remeasured HIPASS data. Distances ($d$) to the galaxies are Hubble flow distances except for IC 1993, where we adopt the distance of the Fornax cluster. $DEF$ is the \HI-deficiency parameter \citep{Haynes1983}.}
\begin{tabular}{l c c c c c c c c }
\hline
Galaxy name& R.A.  & Dec. & Morphology & Environment & $d$ & M$_{\star}$ & M$_{HI}$ & DEF \\
&[hh mm ss.s] & [dd mm ss.s] &  && [Mpc]  & [$10^{9}$ \Msun] & [$10^{8}$ \Msun] & \\
\hline
IC 1993 & 03 47 04.8 & -33 42 35 & Sb & Fornax cluster & 19.3  &  6.3  & 2.0 & 1.09 \\
NGC 1515 & 04 04 02.7 & -54 06 00 & Sbc & Dorado group & 13.8 & 31.6 & 5.6 & 1.01  \\
NGC 6808 & 19 43 54.0 & -70 38 00 & Sab/pec & NGC 6846 group & 45.8 & 31.6 & 25.3 & 0.5 \\
NGC 1473 &  03 47 26.3 & -68 13 14 & Im & group member & 16.8  &  2.0 & 3.8 & 0.99 \\
NGC 6699 & 18 52 02.0 & -57 19 15 & Sbc & group member & 47.0  & 39.8 & 19.8 & 0.76 \\
ESO 009- G 010 & 17 39 31.6 & -85 18 37  & Sbc? & in a triplet & 31.8  &  15.8 & 18.1 & 0.76 \\
\hline
\end{tabular}
\label{tab:properties_summary}
\end{table*}

\subsection{Observations with ATCA}

High resolution synthesis \HI\ line observations for the six \HI-deficient galaxies were obtained with the Australia Telescope Compact Array (ATCA) using a single pointing in two antenna configurations $\sim$12 hours each. Observations with the 750C array were carried out in January-February 2013 and observations with the 1.5B array in January 2014. Details of the observations are given in Table~\ref{tab:observations}. The ATCA is a radio-interferometer consisting of six 22 m dishes, creating 15 baselines in a single configuration. While five antennas (CA01-CA05) are movable along a 3 km long east-west track (and a 214 m long north-south spur), one antenna (CA06) is fixed at a distance of 3 km from the end of the track creating the longest baselines. In this work we do not use data from antenna six (CA06).   

Using only 5 antennas our combined data consists of 20 baselines ranging from 30 m to 1.3 km. We observed each of the sample galaxies for 10 to 15 hours in each array, which resulted in good \textit{uv} coverage for imaging. Our observing strategy was 55 minutes on source and 5 minutes on a phase calibrator (see Table~\ref{tab:observations}) in each hour and 10 minutes at the start or at the end of the observation on the ATCA flux calibrator PKS 1934-638. We used the 1M-0.5k correlator configuration on the new Compact Array Broad-band Backend (CABB; \citealt{Wilson2011}) with a 8.5 MHz wide zoom band\footnote{The 8.5 MHz wide zoom band consists of 16 concatenated 1 MHz zoom channels, each with 2048 channels overlapped by 50 \% to obtain a flat bandpass.} divided into 17408 channels. This gives a velocity resolution of 0.103 \kms. The zoom bands were centred at the frequencies given in Table~\ref{tab:observations}.

\begin{figure*}
	\subfigure{\includegraphics[width=8.8cm]{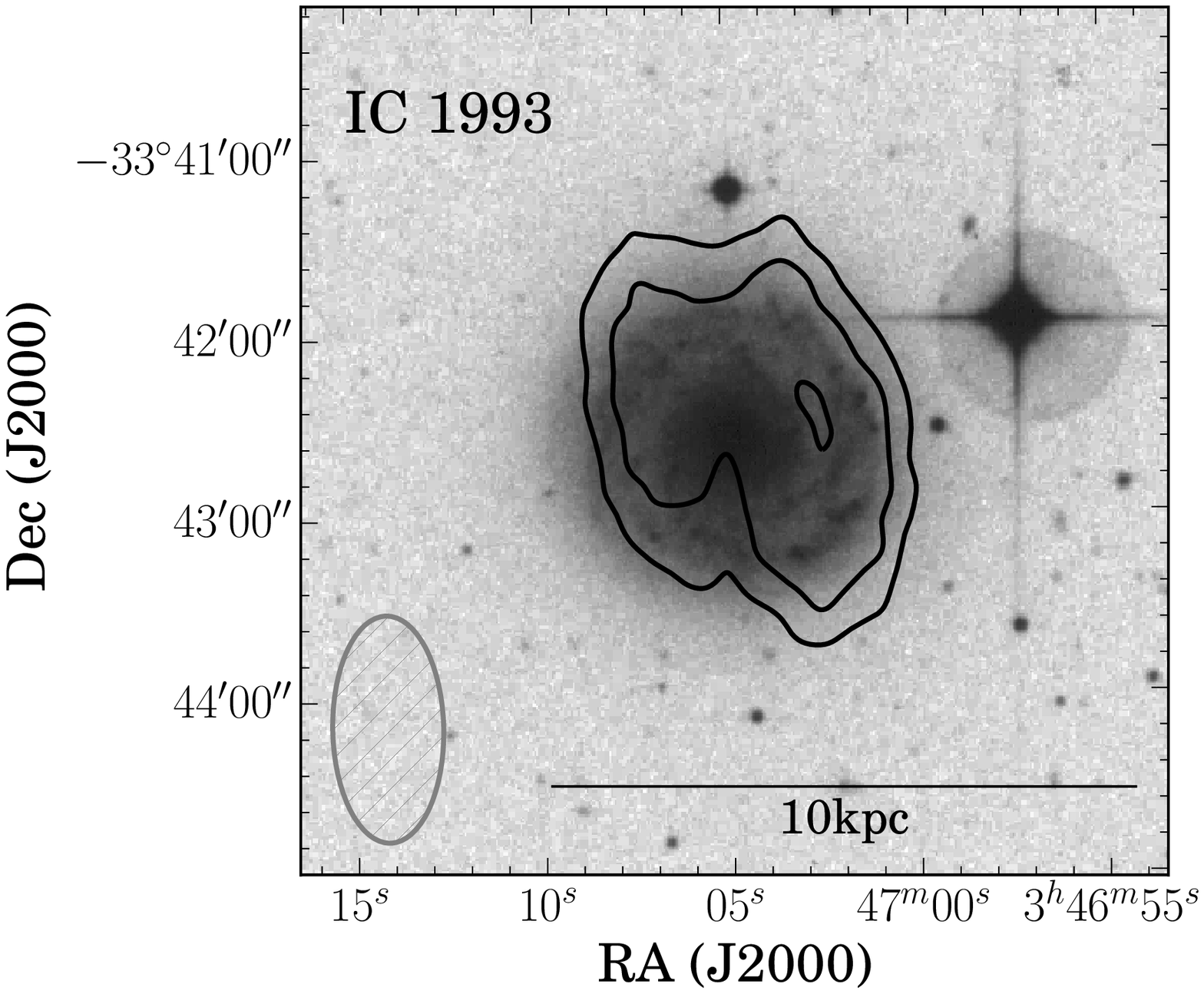}} 
	\subfigure{\includegraphics[width=8.8cm]{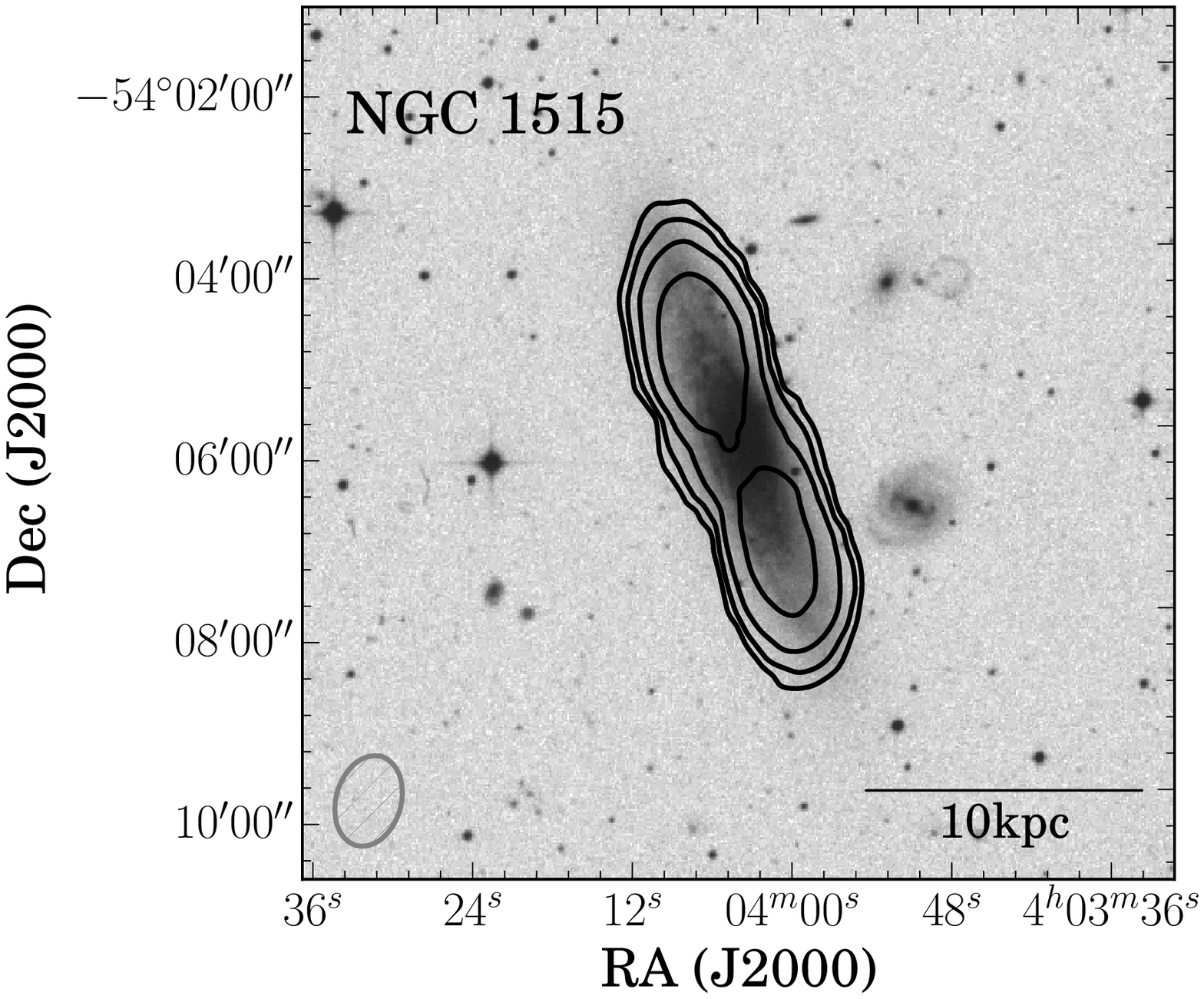}} 
	\subfigure{\includegraphics[width=8.8cm]{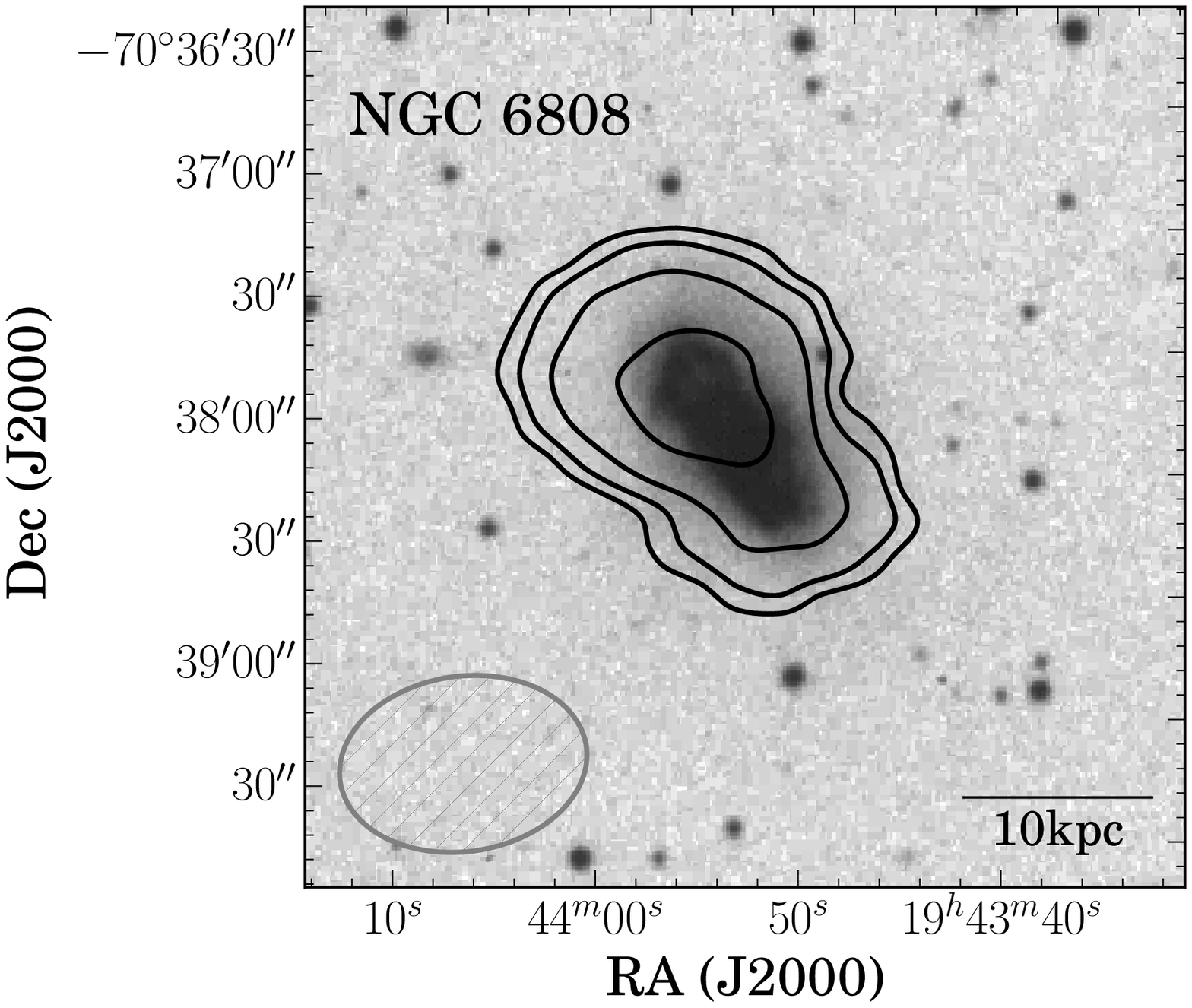}} 
	\subfigure{\includegraphics[width=8.8cm]{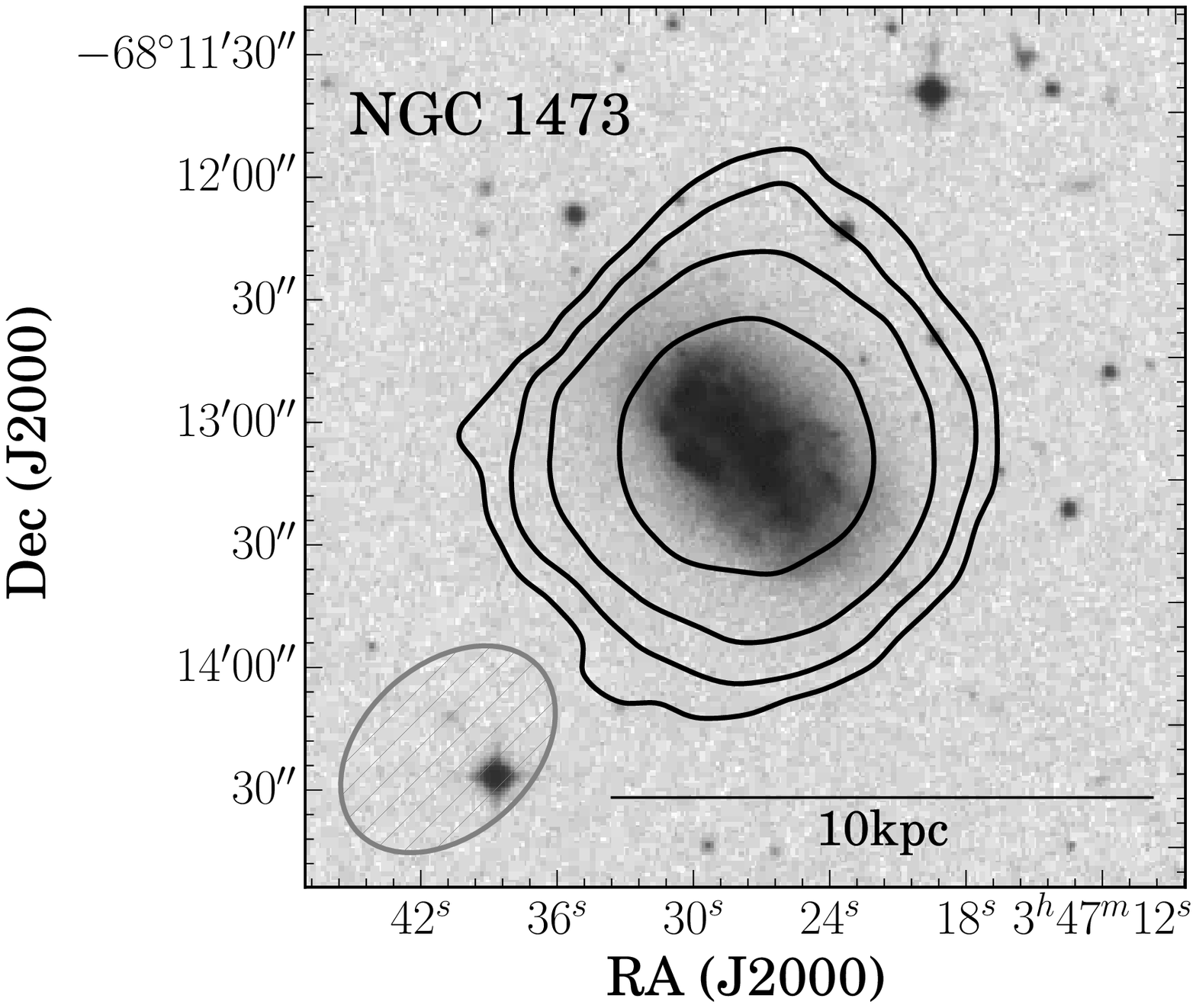}} 
	\subfigure{\includegraphics[width=8.8cm]{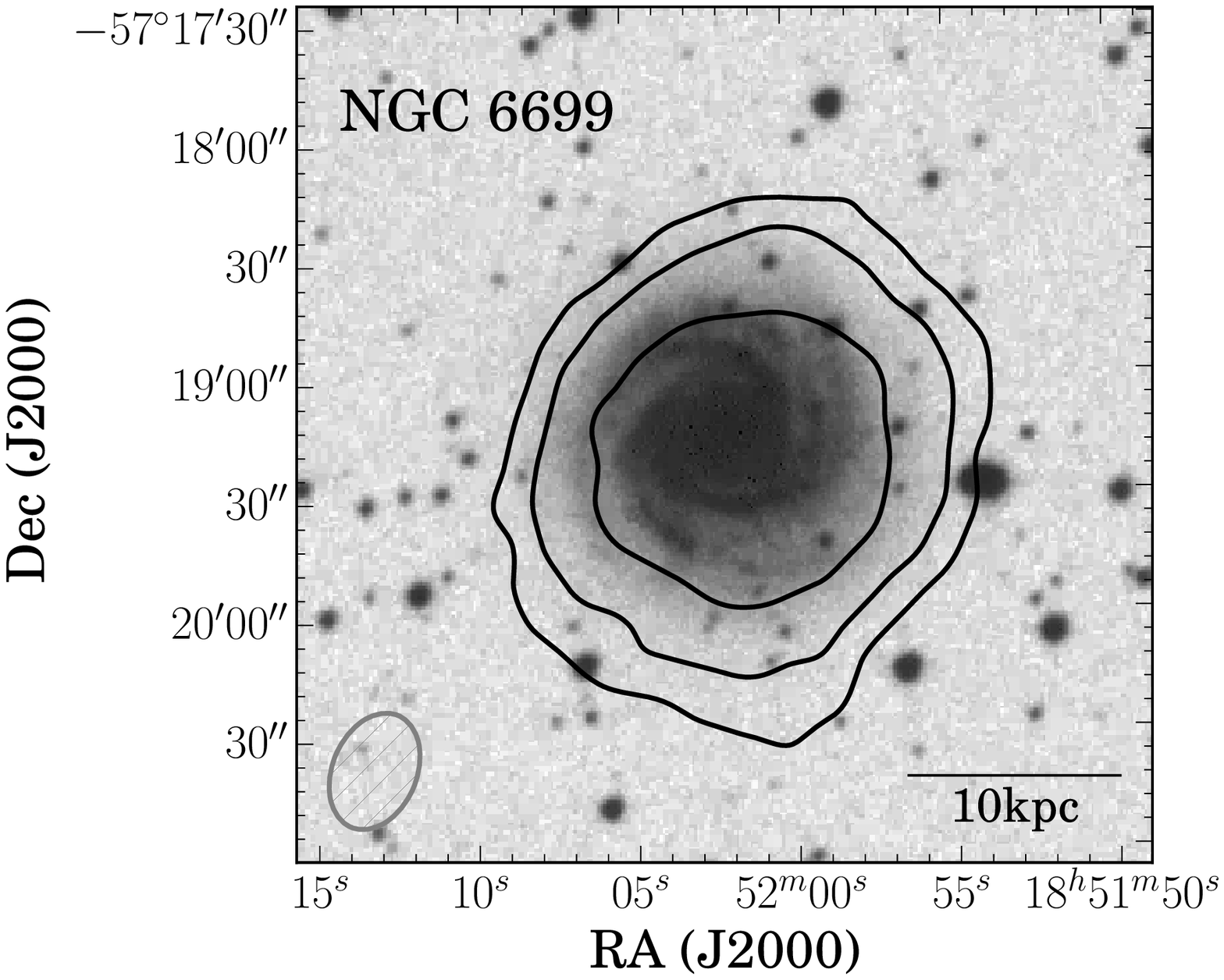}} 
	\subfigure{\includegraphics[width=8.8cm]{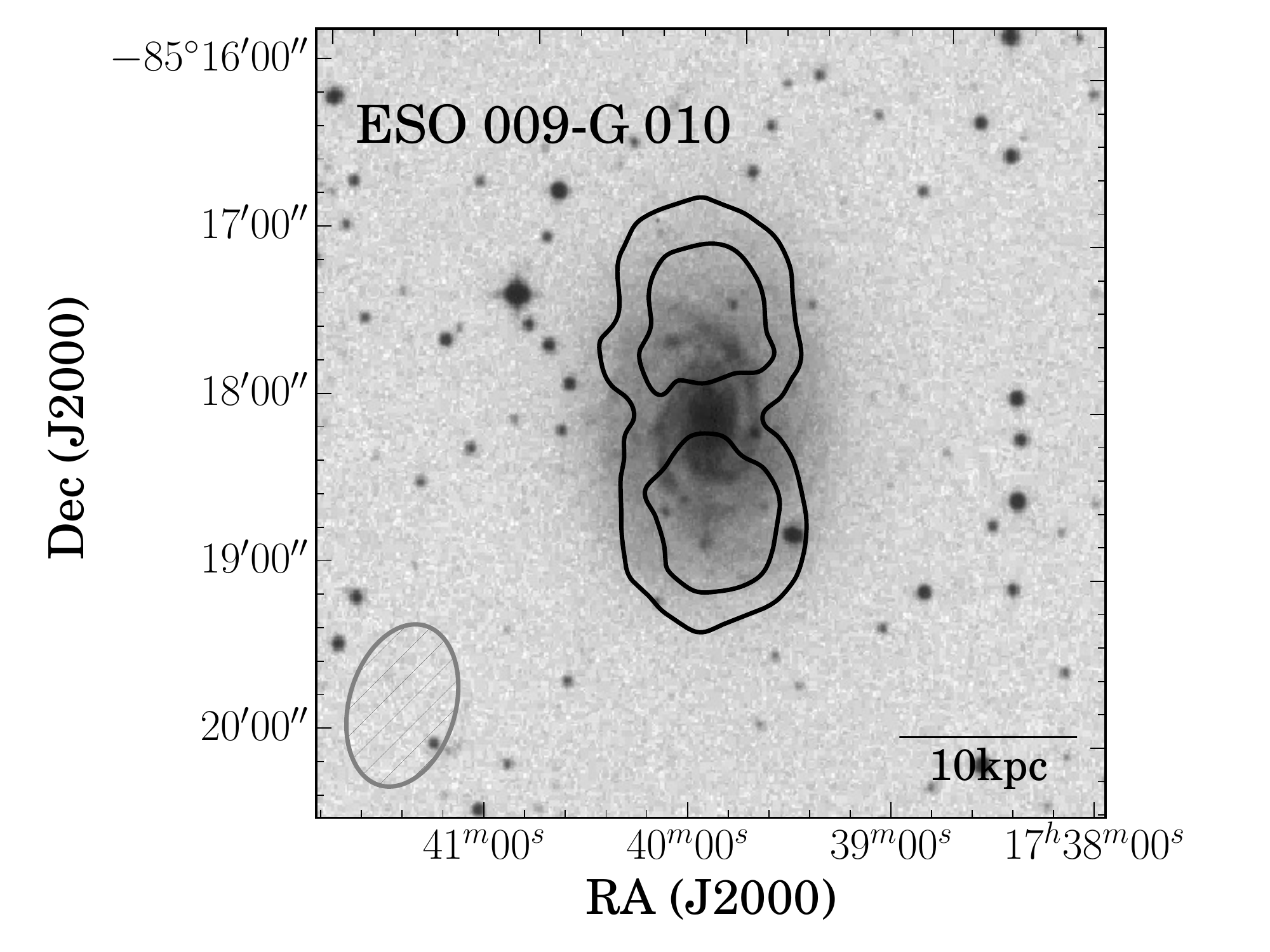}} 
	\caption{\HI\ distribution contours from the combined ATCA data overlaid onto optical B-band images from SuperCOSMOS (UKST). Contour levels are 8, 16, 32, 64 $\times\ 10^{19}$cm$^{-2}$, except for IC 1993 where contour levels are 4, 8, 16 $\times\ 10^{19}$cm$^{-2}$. We show the synthesised beam size in the lower left corner and the size scale (10 kpc) in the lower right corner of each image.}
	\label{fig:HI-distribution}
\end{figure*}

\begin{figure*}
	\subfigure{\includegraphics[width=5.6cm]{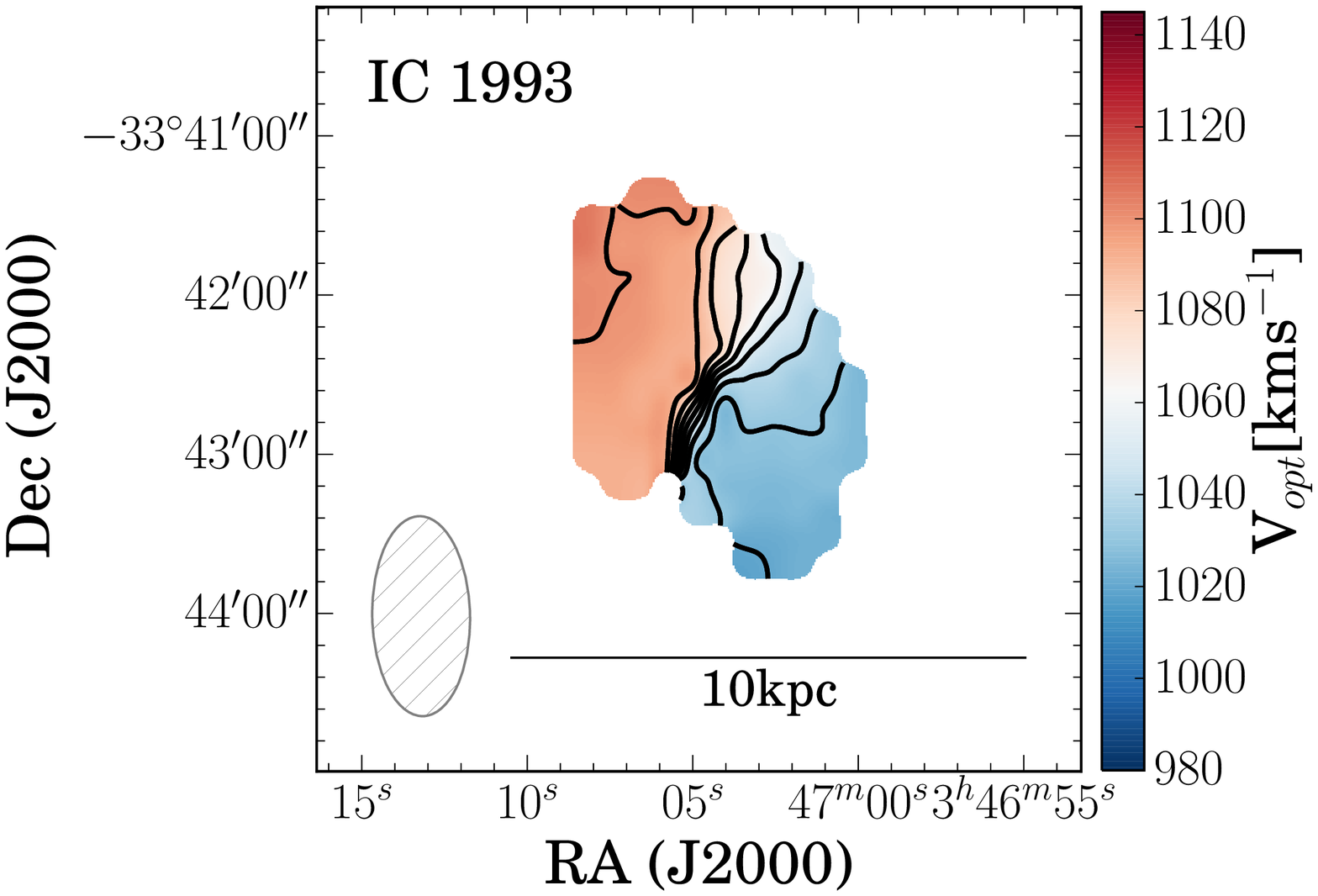}} 
	\subfigure{\includegraphics[width=5.6cm]{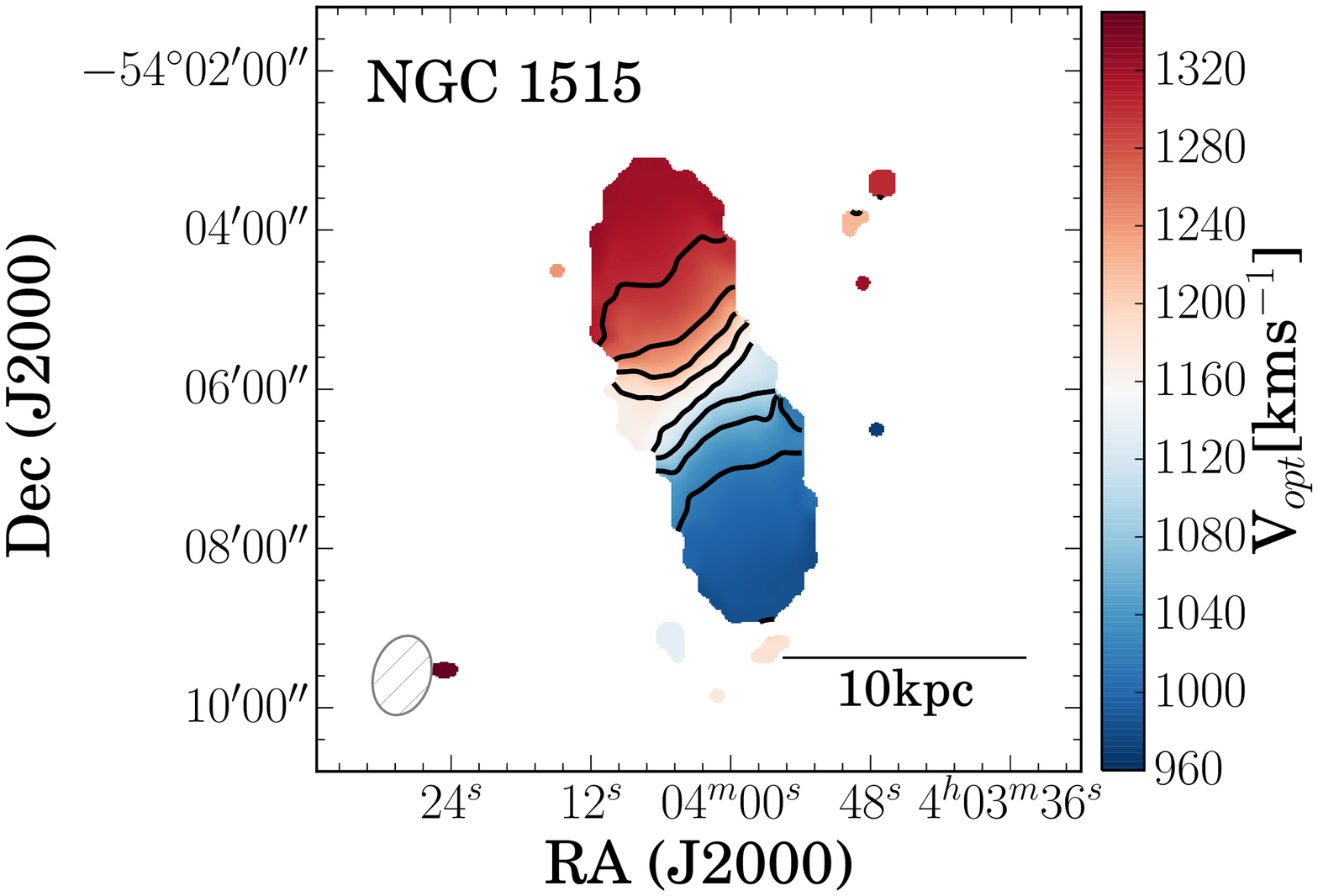}} 
	\subfigure{\includegraphics[width=5.6cm]{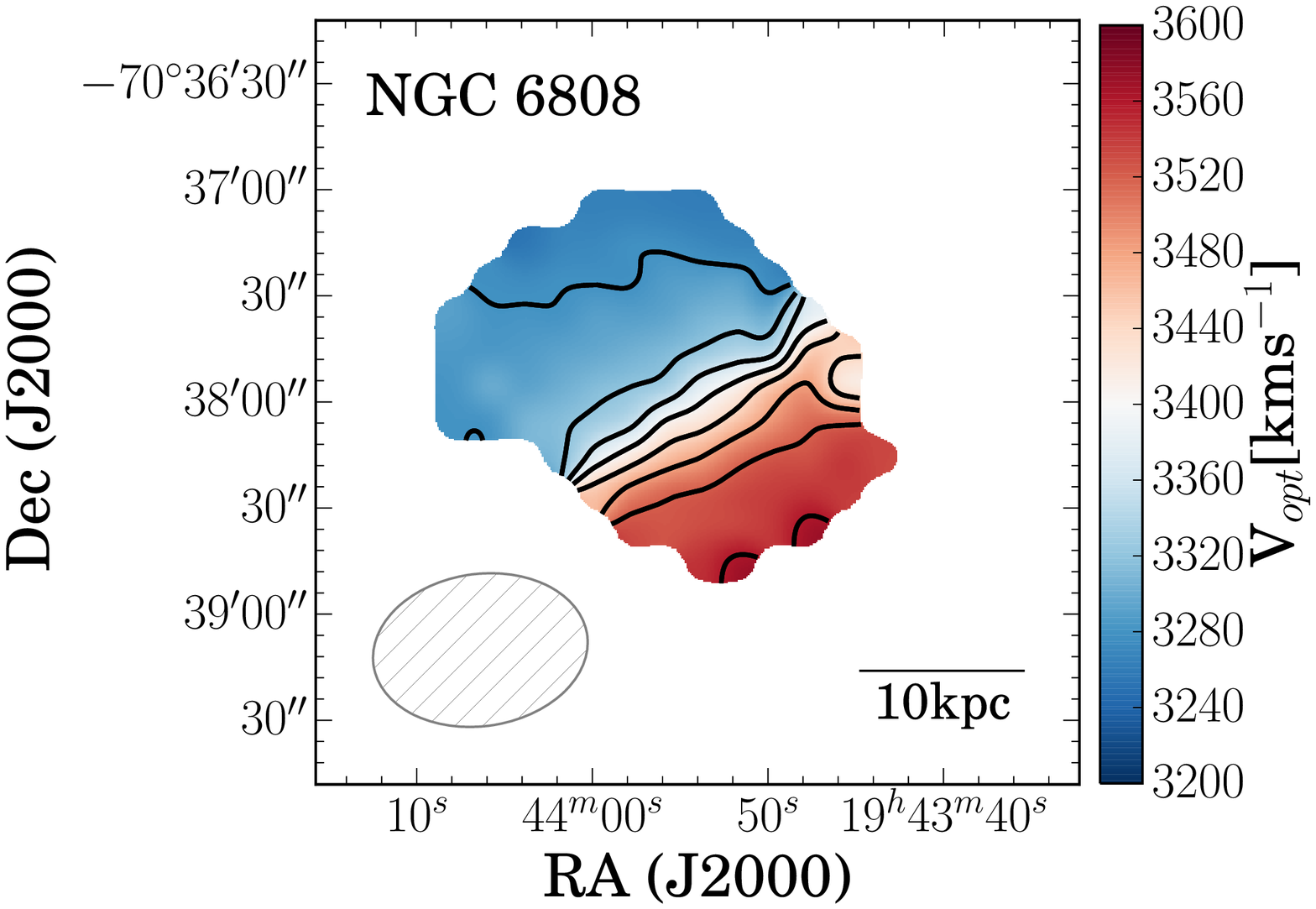}} 
	\subfigure{\includegraphics[width=5.6cm]{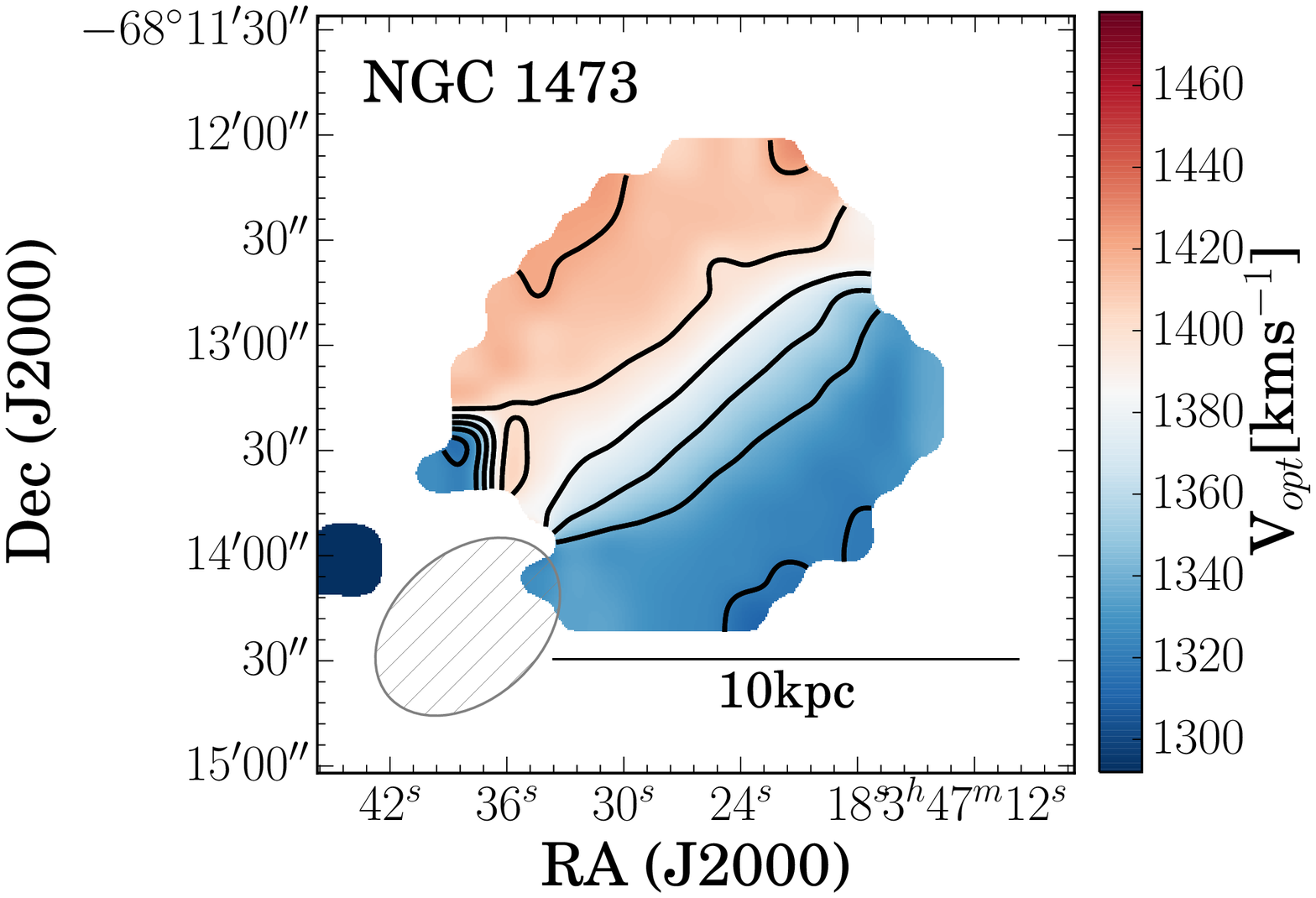}} 
	\subfigure{\includegraphics[width=5.6cm]{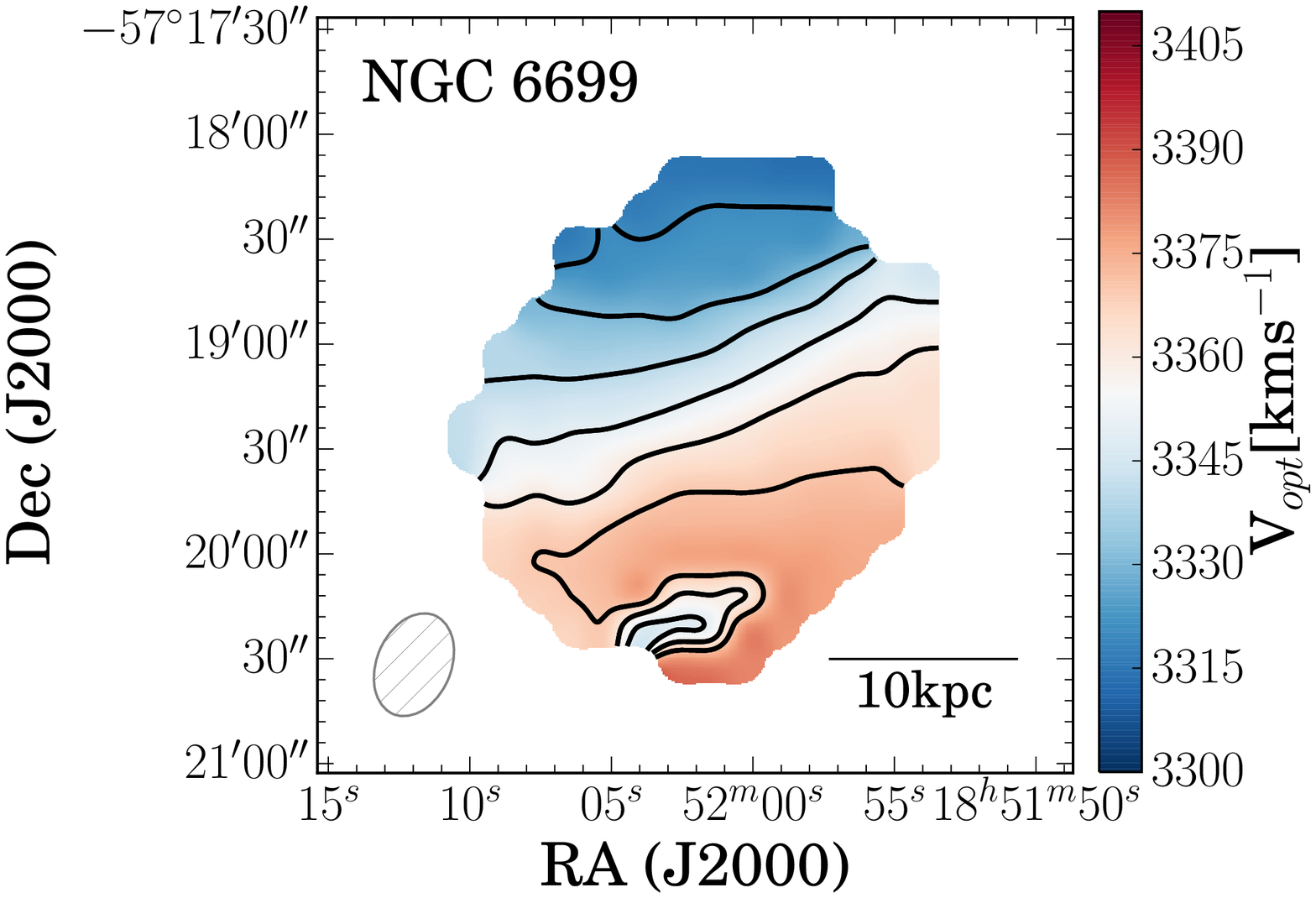}} 
	\subfigure{\includegraphics[width=5.6cm]{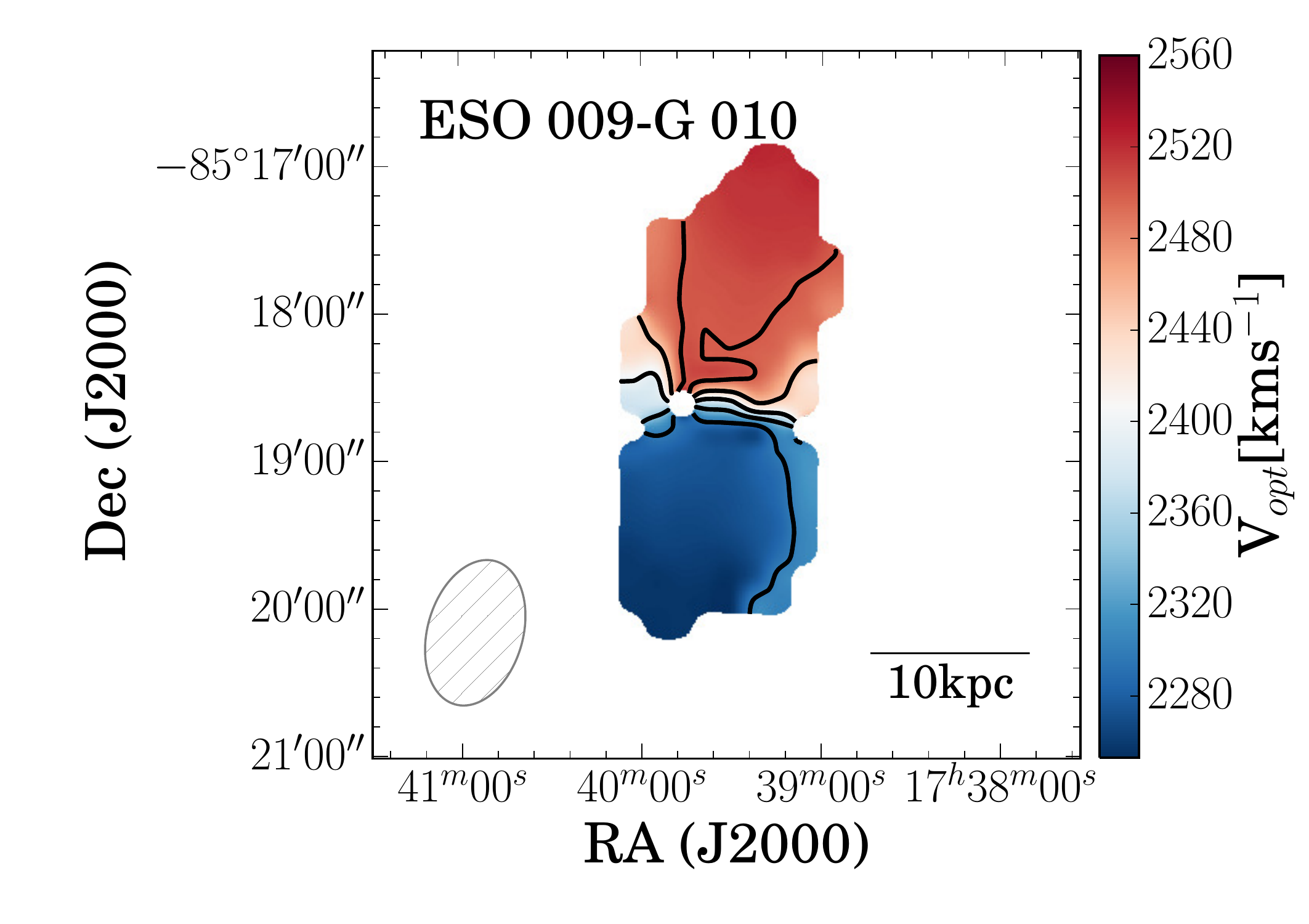}} 
	\caption{\HI\ velocity fields of the sample galaxies. We show the same area and layout as Fig.~\ref{fig:HI-distribution}. Dashed lines are velocity contours smaller than the systemic velocity of the galaxies and solid lines are velocity contours larger than the systemic velocity of the galaxies. Velocity contours are spaced 10 \kms\ for IC 1993 and NGC 6699, 20 \kms\ for NGC 6808 and NGC 1473 and 40 \kms\ for NGC 1515 and ESO 009-G 010. }
	\label{fig:velocity-distribution}
\end{figure*}

\begin{figure*}
	\subfigure{\includegraphics[width=5.5cm]{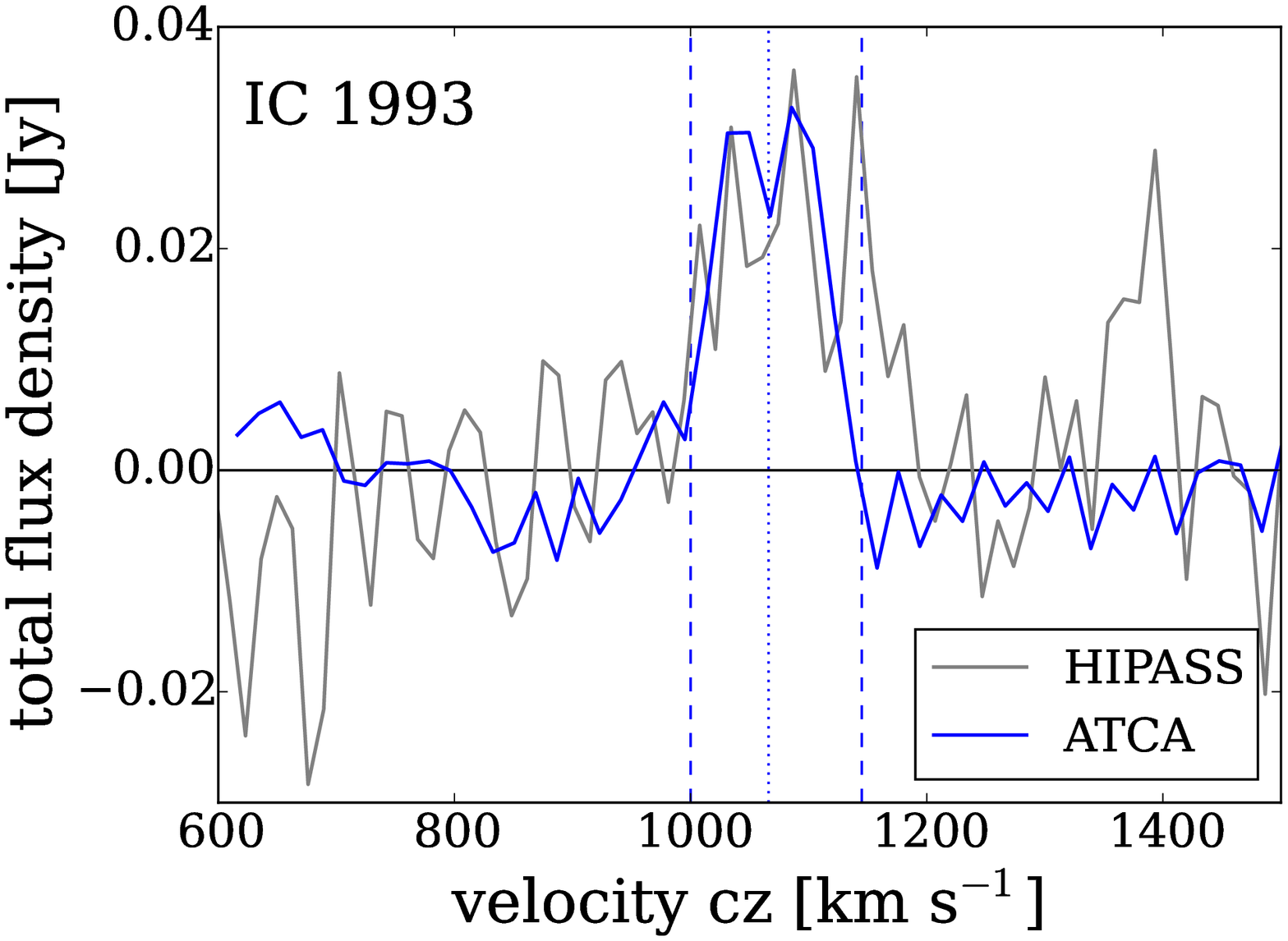}} 
	\subfigure{\includegraphics[width=5.5cm]{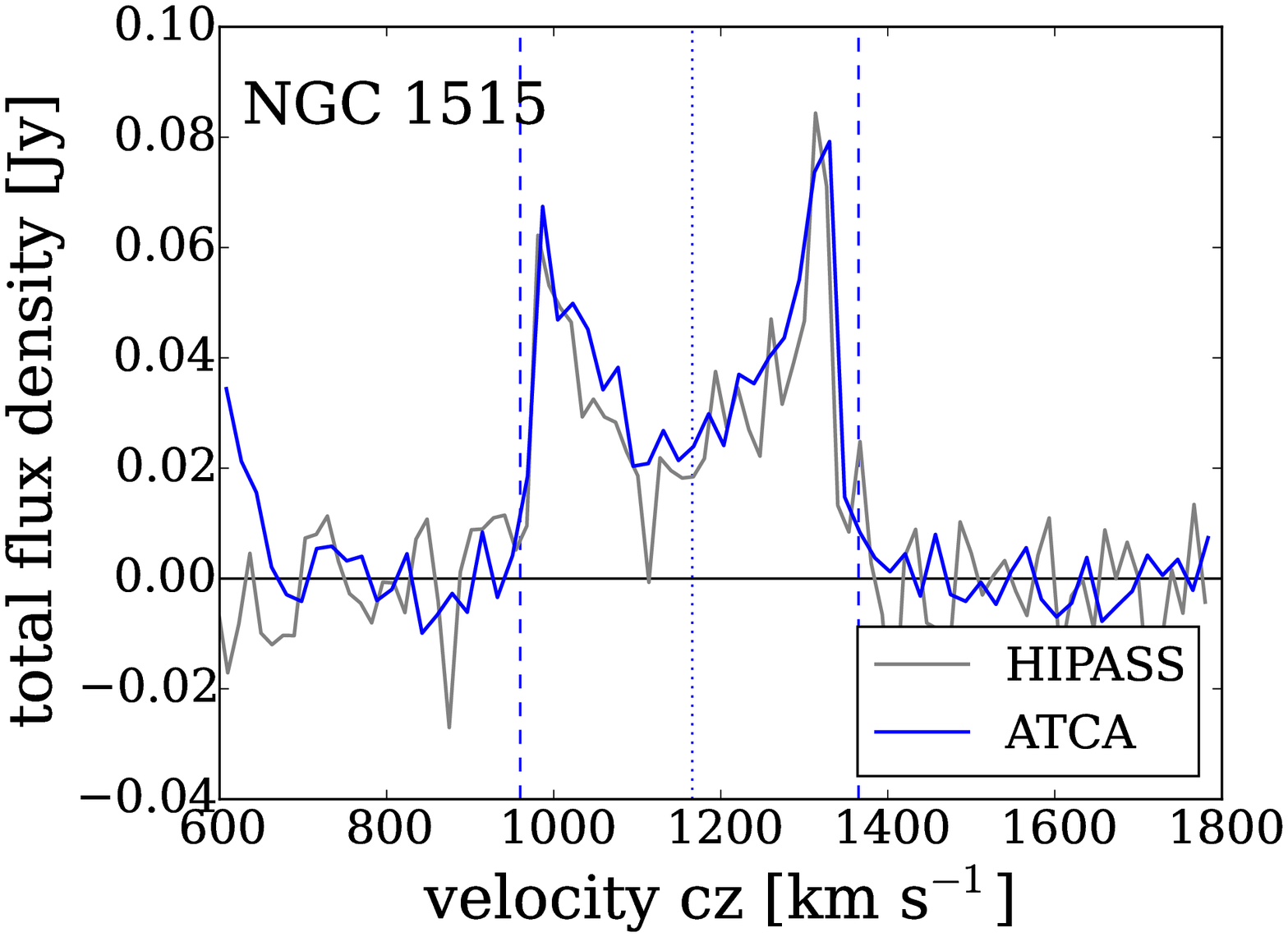}} 
	\subfigure{\includegraphics[width=5.5cm]{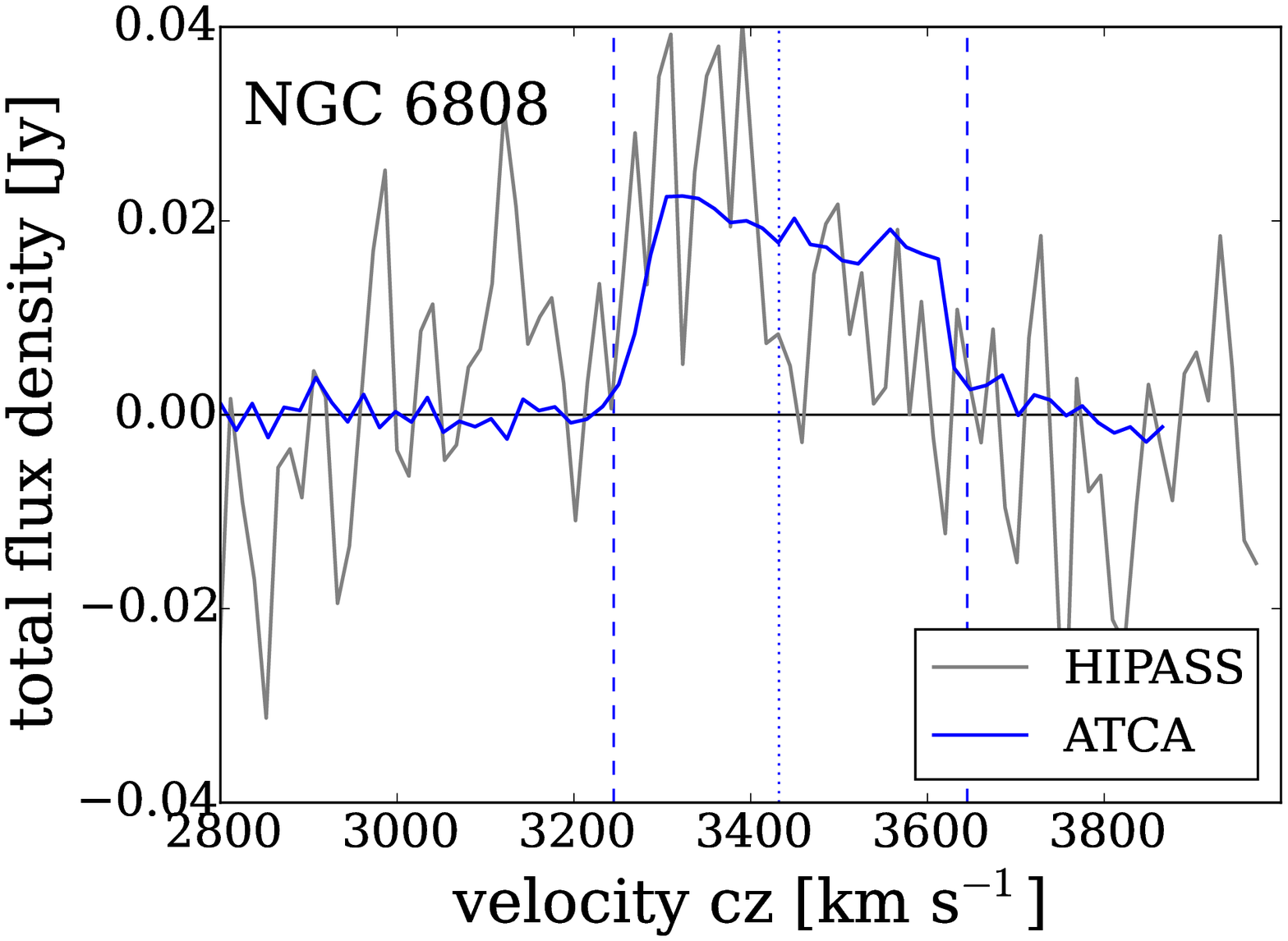}} 
	\subfigure{\includegraphics[width=5.5cm]{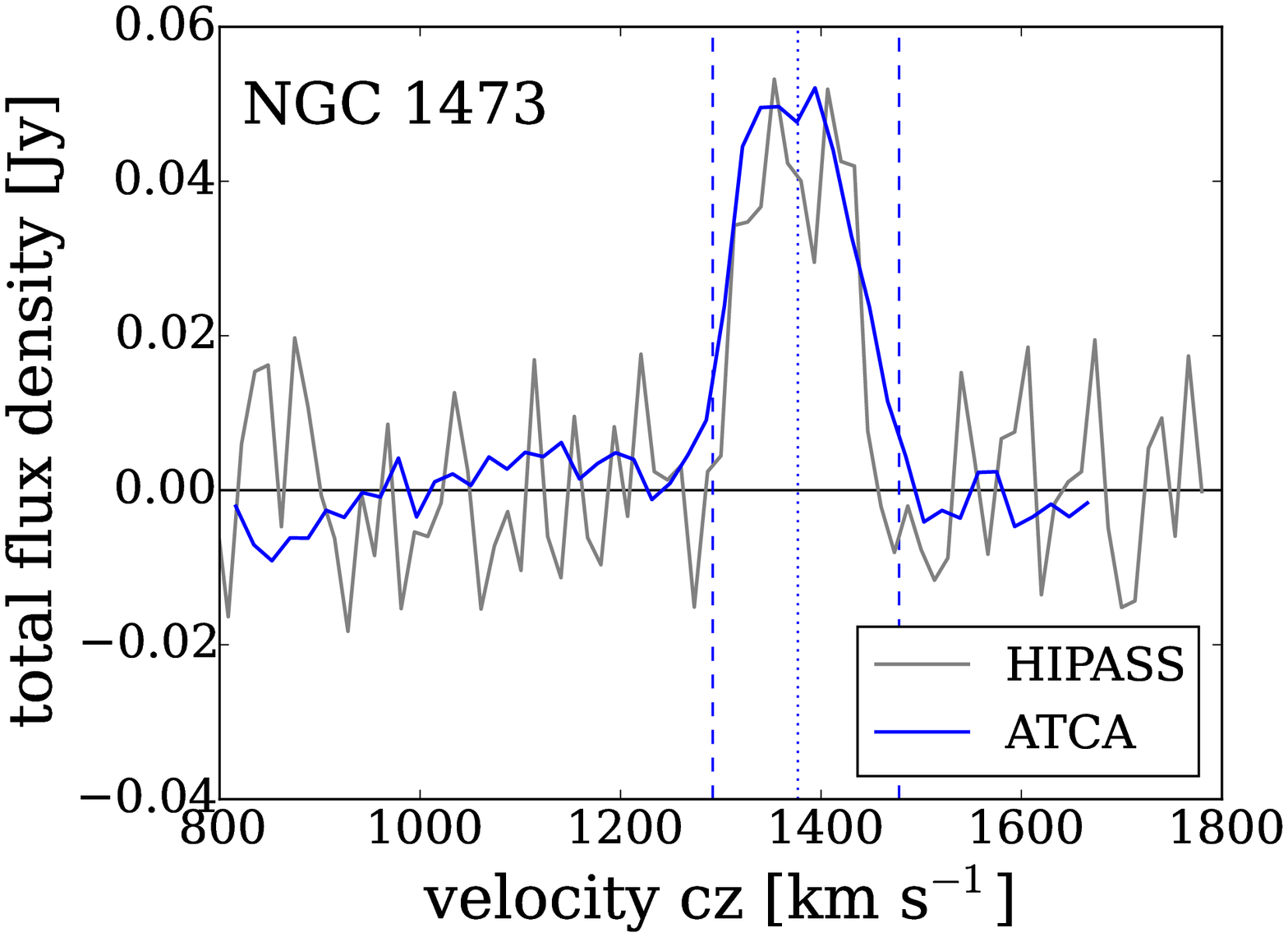}} 
	\subfigure{\includegraphics[width=5.5cm]{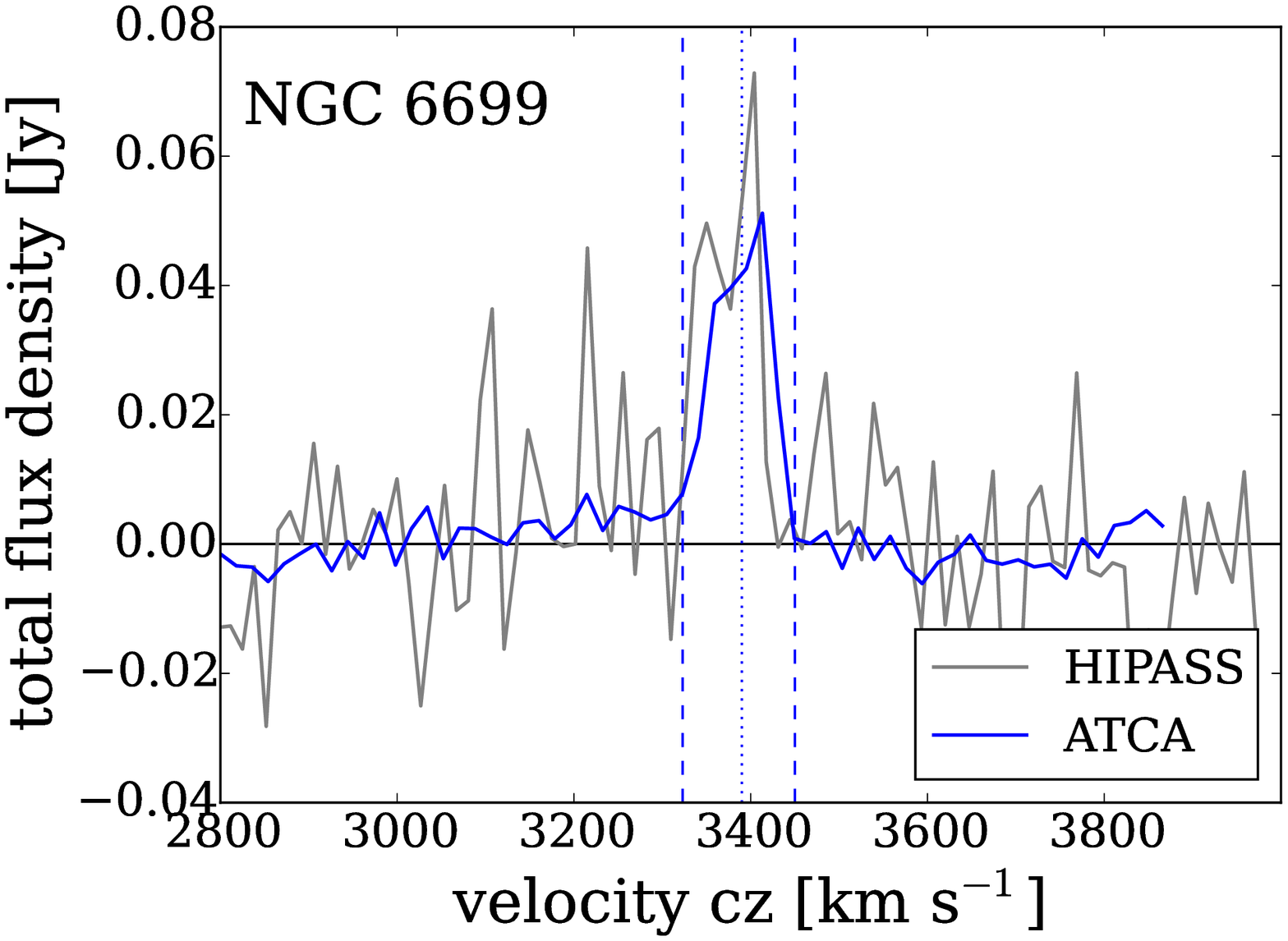}} 
	\subfigure{\includegraphics[width=5.5cm]{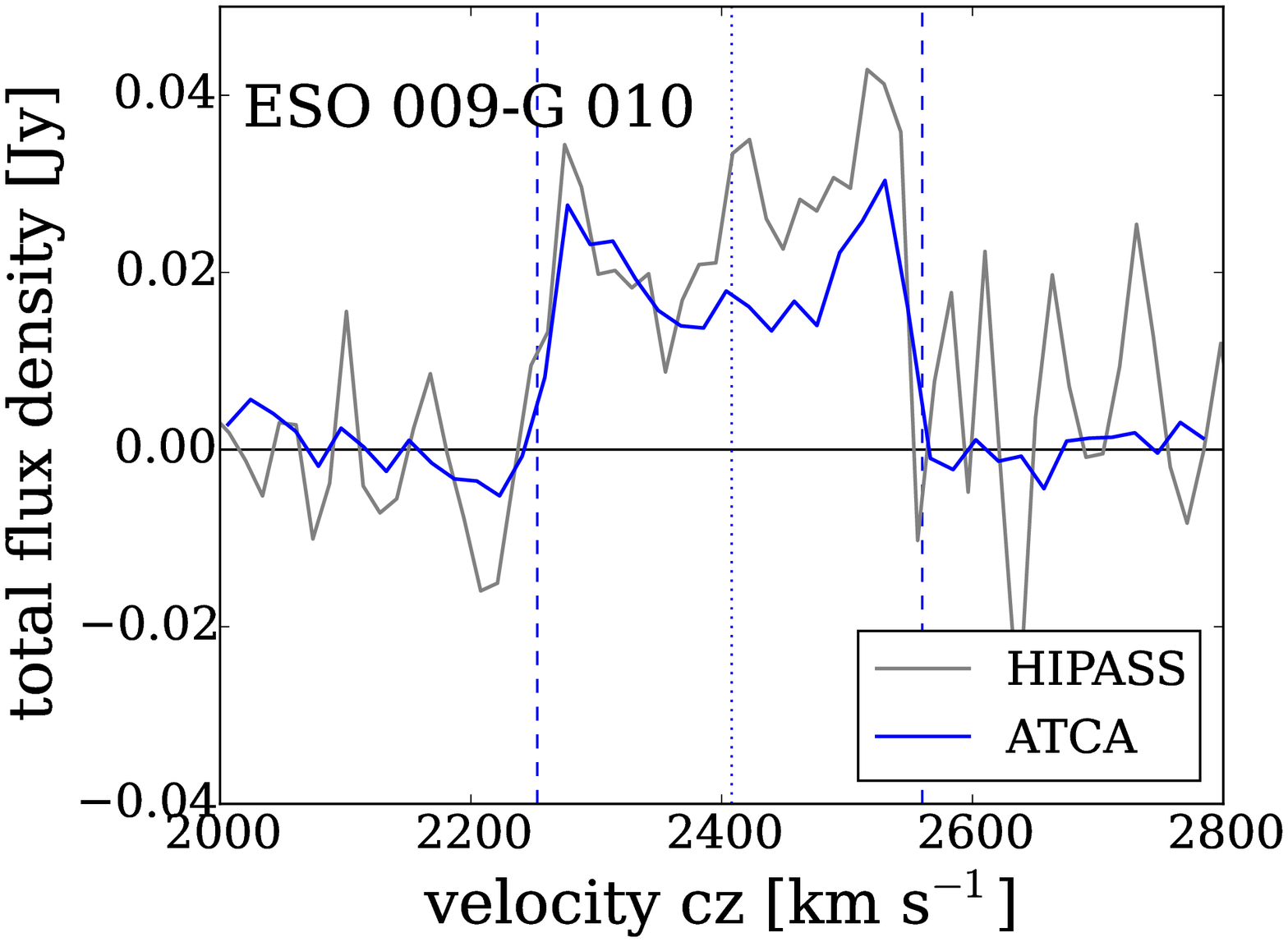}} 
	\caption{Comparison of HIPASS and ATCA baseline subtracted \HI\ profiles. A first order baseline fit is applied to both profiles and the he ATCA profile is smoothed to 18 \kms\ to match the resolution of HIPASS. Vertical dashed lines show where we integrate the \HI\ profile. The blue dotted line shows the measured systemic velocity of the galaxies.}
	\label{fig:HI-profiles}
\end{figure*}

Data reduction was carried out with the {\sc Miriad} software package \citep{Miriad} using standard procedures. After calibration the data was split into \HI\ line data and narrow band 20 cm radio continuum emission using a first order fit to the line-free channels. \HI\ cubes were made using natural and robust weighting with 3 \kms\ velocity steps. We give the measured HIPASS and ATCA \HI\ properties of the sample (measured in the natural weighted data cubes) in Table.~\ref{tab:HI-properties}. We present the \HI\ distribution of the sample in Fig.~\ref{fig:HI-distribution} and \ref{fig:HI-distribution_highres}. The velocity fields are presented in Fig.~\ref{fig:velocity-distribution} and the ATCA and HIPASS \HI\ line profiles in Fig.~\ref{fig:HI-profiles}. The \HI\ line profiles were extracted from the HIPASS and the ATCA data cubes at the same position and over the same velocity range. 

After deriving the \HI\ properties from the ATCA data we remeasured the \HI\ properties from the HIPASS data cubes using the position and velocity width information from the ATCA data. In Table~\ref{tab:HI-properties} we present the remeasured HIPASS values (black crosses in Fig.~\ref{fig:scaling-relation}). This step is important, because the \HI\ masses in HICAT were measured with an automated algorithm with no prior knowledge of the properties of the \HI\ profile. The catalogue measurements are the most reliable (95\% confidence level) for sources with an integrated flux $> 5$ Jy \kms. Given that \HI-deficient sources tend to be faint, occasionally the low signal to noise level or strong RFI spikes might cause errors in the flux measurement. Apart from IC 1993 and NGC 6808 the \HI\ properties from HICAT are in good agreement with our remeasured values. IC 1993 is a marginal detection in HIPASS and has a very noisy spectrum, where it is difficult to determine the correct width of the \HI\ profile only based on HIPASS data. There is archival \HI\ data available for IC 1993 measured with the Nan\c{c}ay telescope \citep{Theureau2005}. The integrated flux (1.57 $\pm$ 0.53 Jy \kms) and the \HI\ profile of the Nan\c{c}ay data are in good agreement with our ATCA measurements. In the case of NGC 6808 the HIPASS data has a large negative noise spike in the middle of the \HI\ profile. This resulted in only including the blue side of the galaxy spectra in HICAT, which is also evident from the considerably smaller $w_{50}$ and $w_{20}$ line widths in HICAT. The measured \HI\ flux in HICAT is only about half of the remeasured flux from the HIPASS data cubes. With the new flux measurement NGC 6808 is not as \HI-deficient as initially categorized, but with 3 times less \HI\ than expected it is still \HI-deficient. NGC 6808 remains in the sample, because it has still a much lower gas content than a late-type galaxy of similar brightness and it is located in a relatively low density environment. 

We detected 1.4 GHz continuum emission from four of the sample galaxies. The 1.4 GHz narrow band continuum maps of the ATCA observations are presented in Fig.~\ref{fig:continuum}. The rms of the continuum maps is between 0.002-0.003 Jy beam$^{-1}$. The measured continuum fluxes and upper limits for the non detections are presented in Table~\ref{tab:star-formation}.

\begin{table*}
\caption{Summary of ATCA observations. $^{\star}$ The sensitivity of the ATCA observations is calculated for the natural weighted data cubes at a 5$\sigma$ rms level over the whole line width of each galaxy.}
\begin{tabular}{l c c c c c c}
\hline
Galaxy name& integration time & central frequency & phase calibrator & Synthesised beam & rms noise & Sensitivity$^{\star}$\\
& 750C, 1.5B [min] & [MHz] &  & [''] & [mJy/beam] & [$10^{19}$ cm$^{-2}$] \\
\hline
IC 1993 & 387, 594 & 1414.75 & PKS 0332-403 & 75 x 37 & 1.4 & 2.7\\
NGC 1515 &  571, 658 & 1414.75 & PKS 0302-623 & 61 x 43 & 1.3 & 6.6 \\
NGC 6808 & 860, 750 & 1404.75 & PKS 1934-638, PKS 2146-783 & 61 x 43 & 1.9 & 9.8 \\
NGC 1473 & 673, 709 & 1413.75 & PKS 0302-623 & 60 x 42 & 1.3 & 3.5\\
NGC 6699 & 542, 613 & 1403.75 & PKS 1934-638 & 60 x 42 & 1.4 & 1.9\\
ESO 009- G 010 & 788, 887 & 1408.75 & PKS 2146-783 & 60 x 39 & 1.2 & 5.2\\
\hline
\end{tabular}
\label{tab:observations}
\end{table*}

\begin{table*}
\caption{\HI\ properties of the sample measured from ATCA observations and the HIPASS data cubes. We present here the re-measured \HI\ properties from the HIPASS data cubes. $\star$ First flux ratio is ATCA/HIPASS, second is ATCA 1.5k/ATCA 750m.}
\begin{tabular}{l l c c c c c c c c}
\hline
Telescope & Galaxy name  & $S_{int}$ & $S_{peak}$ & $v$ & $w_{50}$ & $w_{20}$ & M$_{HIobs}$ & \textit{DEF} & flux ratio$\star$ \\
&  & [Jy km s$^{-1}$] & [mJy] & [km s$^{-1}$] & [km s$^{-1}$] & [km s$^{-1}$] & [$10^8$M$_{\odot}$] & & $\frac{ATCA}{HIPASS}$ \\
\hline
Parkes  & HIPASS J0347-33 & 2.3 $\pm$ 0.3 & 33 $\pm$ 1& 1072 $\pm$ 1& 97 $\pm$ 3& 114 $\pm$ 4& 2.0 $\pm$ 0.4 & 1.09 $\pm$ 0.08 & 0.7 $\pm 0.04$ \\
ATCA & IC1993 & 1.6 $\pm$ 0.1 & 29 $\pm$ 1& 1066 $\pm$ 1& 90 $\pm$ 1& 106 $\pm$ 2& 1.4 $\pm$ 0.1 & 1.24 $\pm$ 0.03 & 1.0 $\pm 0.08$\\
\hline
Parkes  & HIPASS J0404-54 & 11.6 $\pm$ 1.1 & 82 $\pm$ 1& 1173 $\pm$ 1& 357 $\pm$ 1& 369 $\pm$ 2& 5.2 $\pm$ 1.7 & 1.01 $\pm$ 0.12 & 0.99 $\pm 0.08$\\
ATCA  & NGC1515 & 11.5 $\pm$ 0.1 & 91 $\pm$ 1& 1166 $\pm$ 1 & 356 $\pm$ 1 & 371 $\pm$ 1& 5.2 $\pm$ 0.2 & 1.02 $\pm$ 0.01 & 0.64 $\pm 0.01$\\
\hline
Parkes  & HIPASS J1944-70 & 5.1 $\pm$ 0.9 & 37 $\pm$ 1& 3436 $\pm$ 2& 306 $\pm$ 4& 340 $\pm$ 6& 25.3 $\pm$ 10.1 & 0.5 $\pm$ 0.15 & 0.78 $\pm 0.08$\\
ATCA  & NGC6808 & 4.0 $\pm$ 0.2 & 20 $\pm$ 2& 3432 $\pm$ 2& 334 $\pm$ 4& 367 $\pm$ 6& 19.6 $\pm$ 2.1 & 0.61 $\pm$ 0.04 & 1.0 $\pm 0.09$\\
\hline
Parkes  & HIPASS J0347-68 & 5.7 $\pm$ 0.2 & 54 $\pm$ 1& 1392 $\pm$ 1& 130 $\pm$ 2& 144 $\pm$ 2& 3.8 $\pm$ 0.3 & 0.99 $\pm$ 0.04 & 0.75 $\pm 0.04$ \\
ATCA  & NGC1473 & 4.3 $\pm$ 0.1 & 41 $\pm$ 1& 1377 $\pm$ 1& 132 $\pm$ 1& 171 $\pm$ 2& 2.8 $\pm$ 0.1 & 1.11 $\pm$ 0.02 & 0.64 $\pm 0.01$\\
\hline
Parkes  & HIPASS J1851-57 & 3.8 $\pm$ 0.3 & 70 $\pm$ 1& 3391 $\pm$ 1& 76 $\pm$ 1& 90 $\pm$ 2& 19.8 $\pm$ 3.1 & 0.76 $\pm$ 0.06 & 0.66 $\pm 0.02$\\
ATCA  & NGC6699 & 2.5 $\pm$ 0.1 & 47 $\pm$ 1& 3390 $\pm$ 1& 66 $\pm$ 1& 107 $\pm$ 2& 13.2 $\pm$ 0.6 & 0.94 $\pm$ 0.02 & 0.63 $\pm 0.01$\\
\hline
Parkes  & HIPASS J1740-85 & 7.6 $\pm$ 0.5 & 42 $\pm$ 1& 2419 $\pm$ 1& 287 $\pm$ 3& 309 $\pm$ 4& 18.1 $\pm$ 3.3 & 0.76 $\pm$ 0.07 & 0.42 $\pm 0.01$\\
ATCA  & ESO009-G010 & 3.2 $\pm$ 0.1 & 26 $\pm$ 1& 2408 $\pm$ 1& 277 $\pm$ 1& 291 $\pm$ 2& 7.7 $\pm$ 0.4 & 1.13 $\pm$ 0.02 & 0.87 $\pm 0.03$\\
 \hline
 \end{tabular}
 \label{tab:HI-properties}
 \end{table*}

\subsection{Optical, Infrared and UV data}

Archival optical, UV and infrared data is available for the sample. We obtain 2nd Digital Sky Survey (DSS2) Blue \footnote{http://skyview.gsfc.nasa.gov} images and SuperCOSMOS (UKST) Blue \footnote{http://www-wfau.roe.ac.uk} images to compare the \HI\ morphology of the galaxies to their optical morphology and size. \HI\ contour maps overlaid on SuperCOSMOS B-band images are presented in Figs.~\ref{fig:HI-distribution} and \ref{fig:HI-distribution_highres}. The Carnegie Irvine Galaxy Survey (CGS, \citealt{Ho2011}) imaged 4 of the 6 sample galaxies (IC 1993, NGC 1515, NGC 6699 and ESO 009-G 010). We adopt from CGS the B-band strength of lopsidedness ($ \langle I_{1}/I_{0}  \rangle _{B}$), which is a parameter based on the Fourier analysis of radial profiles \cite{Li2011}.

IC 1993 has archival Hubble Space Telescope (HST) imaging data available from the Wide Field Planetary Camera 2 (WFPC2) in two wavebands (F814W, F450W). NGC 6699 and NGC 6808 have available HST data from the Near Infrared Camera and Multi Object Spectrometer (NICMOS) Camera 3 (NIC3) in two wavebands (F190N, F160W). HST images were obtained from the ESA Huble Science Archive\footnote{http://archives.esac.esa.int/hst/} and are used to compare the high resolution near infrared morphology to the high resolution \HI\ morphology of the sample.  

We obtain FUV and NUV images from the Galaxy Evolution Explorer (GALEX; \citealt{Martin2005})\footnote{http://galex.stsci.edu/GR6/} All-sky Imaging Survey (AIS). ESO 009-G 010 has only NUV data available. We measure the NUV and FUV fluxes from the GALEX images using elliptical apertures. We correct the UV fluxes for galactic absorption following \cite{GildePaz2007}. The UV data is used to calculate star formation rates for our sample and to compare the UV morphologies with the optical and \HI\ morphologies.

The Wide-field Infared Survey Explorer (WISE, \citealt{WISE}) provides infrared images in 3.4, 4.6, 12 and 22$\mu$m for the whole sky. We obtain WISE images and measure the 22$\mu$m fluxes of the sample using the same elliptical apertures as used for the UV flux measurements. We correct the 22 $\mu$m luminosities following \cite{Jarrett2003} and \cite{Ciesla2014}. The 22 $\mu$m data is then combined with the GALEX FUV luminosities to calculate star formation rates following \cite{Hao2011} assuming a Kroupa IMF. The calculated star formation rates are presented in Table~\ref{tab:star-formation}. 

NGC 1515, IC 1993 and NGC 1473 are part of The Spitzer Survey of Stellar Structure in Galaxies (S$^{4}$G, \citealt{Sheth2010}) and have high resolution infrared data available. We obtained Spitzer 3.6$\mu$m and WISE images, from the NASA/IPAC Infrared Science Archive \footnote{http://irsa.ipac.caltech.edu}. These images are used to compare the infrared morphology of the galaxies to their \HI\ morphology.

\section{Results}
\label{Results}

Five per cent (94) of the galaxies in the HIPASS parent sample are \HI-deficient with $DEF > 0.6$. The 6 galaxies presented in this paper span a stellar mass range of $10^{9.3}$ - $10^{10.6}$ \Msun, an \HI\ mass range of $10^{8.2}$ - $10^{9.2}$ \Msun\ (Fig.~\ref{fig:scaling-relation}) and optical morphology classes from Sb to irregular. 

\subsection{Morphological comparison}
 
The \HI\ morphology of the sample is characterized by relatively small \HI\ disks and lopsided \HI\ distributions (Fig.~\ref{fig:HI-distribution}). All galaxies have a smaller \HI\ diameter (measured at 1\Msun pc$^{-2}$) than their 25 \textit{B}-mag arcsec$^{-2}$ optical diameter (Table~\ref{tab:diameters}). The ratio of the \HI\ to the optical diameter of the sample is between 0.41 and 0.95, which is smaller than the average for late-type galaxies. The \HI\ disk of a typical late-type galaxy extends about 1.7 - 2 times the size of its stellar disk \citep{Broeils1994, Broeils1997,Verheijen2001}. Asymmetries in the spatial distribution of the \HI\ are visible in the \HI\ distribution maps overlaid on the B-band optical images. The natural weighted \HI\ images (Fig.~\ref{fig:HI-distribution}) show that IC 1993 and NGC 6808 have strongly lopsided \HI\ disks and the \HI\ disk of NGC 6699 is off centre compared to the optical disk. These characteristics are even more pronounced in the higher resolution, robust weighted, \HI\ images (Fig.~\ref{fig:HI-distribution_highres}). Based on the HIPASS \HI\ fluxes and \HI\ profiles, there is indication of significant diffuse \HI\ around 5 of the galaxies. The column density sensitivity of our ATCA observations is $\sim 10^{19}$ cm$^{-2}$ compared to the $\sim 10^{20}$ cm$^{-2}$ sensitivity of HIPASS \footnote{Assuming a 3$\sigma$ detection and using the natural weighted ATCA data cubes.}.

In Fig.~\ref{fig:HI-profiles} we compare the HIPASS and the ATCA \HI\ profiles of the sample. The ATCA data is smoothed to the same (18 \kms) resolution as the HIPASS data. There appears to be extra \HI\ emission in the receding sides of IC 1993 and ESO 009-G 010 and on the approaching side of NGC 6808. This might be indicative of diffuse gas in these galaxies. However, we note that all three galaxies have a noisy HIPASS spectra. The more detailed 3 \kms\ resolution ATCA \HI\ profiles are presented in Fig.~\ref{fig:HI-profiles2}. We measured the asymmetry of the \HI\ profiles by comparing the flux in the approaching and the receding side of the profile following the method presented in \citep{Espada2011} (see Fig.~\ref{fig:HI-profiles} and Tab.~\ref{tab:diameters}). \cite{Espada2011} characterize galaxies as slightly asymmetric and strongly asymmetric if the asymmetry parameter is larger than 1.15 and 1.51 respectively. Based on this classification, we find that half of the sample has a symmetric \HI\ profile, based on the ATCA measurements. NGC 6808 has a strongly asymmetric \HI\ profile and NGC 1473 and NGC 6699 have slightly asymmetric \HI\ profiles. These galaxies also have lopsided \HI\ distributions. The velocity fields of IC 1993 and NGC 6808 are slightly lopsided and ESO 009-G 010 has a peculiar velocity field (see Fig.~\ref{fig:velocity-distribution}). Strong asymmetries of the \HI\ profile and distribution in galaxies can be tracers of past environmental effects, such as tidal interactions or ram pressure stripping (e.g. \citealt{Richter1994, Espada2011}). Starvation is not expected to produce strongly asymmetric \HI\ disks, since this mechanism only cuts off the gas supply of the galaxy but does not otherwise disturb the \HI\ disk. 

Compared to the \HI\ morphology the stellar morphology of the galaxies is mostly regular. The HST image of NGC 6808 (Fig.~\ref{fig:NGC6808-Hubble}) shows that this galaxy has a strongly warped stellar disk, which is a clear indication of a strong tidal interaction. Four sample galaxies have optical lopsidedness measurements from CGS \citep{Li2011}. \cite{Li2011} define strong lopsidedness as $ \langle I_{1}/I_{0}  \rangle _{B} \ge 0.2 $. The average B-band lopsidedness of IC 1993, NGC 6699 and ESO 009-G 010 indicate significantly lopsided stellar disks. Lopsided stellar disks can be the results of tidal encounters. \citep{Mapelli2008} show that galaxy flybys can produce long lived lopsidedness in stellar disks, whereas their simulations do not produce a lopsided stellar disk for ram pressure stripping. However, from observational studies of the Virgo cluster, we know that strong ram pressure stripping can also affect the dust content of galaxies (e.g \citealt{Abramson2014, Cortese2010}) and trigger star formation in the leading side of the galaxy or in the stripped gas of an \HI\ tail (e.g. \citealt{Abramson2011, Kenney2014}).

\subsection{Star formation rates}

We detect 1.4 GHz continuum emission from four sample galaxies, NGC 1515, NGC 6808, ESO009-G 010 and NGC 6699. Following \cite{Murphy2011}, we calculate star formation rates from the 20 cm radio fluxes. We also derive star formation rates based on the FUV and 22 $\mu$m infrared flux following \cite{Hao2011}. 22 $\mu$m fluxes are corrected following \cite{Jarrett2013} and scaled to 24 $\mu$m fluxes following \cite{Ciesla2014}. Radio continuum, UV and infrared fluxes and the calculated star formation rates are presented in Table~\ref{tab:star-formation}. Except NGC 6808 and NGC 6699, all sample galaxies have moderate star formation rates, between 0.2 and 0.7 \Msun\ yr$^{-1}$, based on their UV emission.  NGC 6808 is a starburst galaxy \citep{Coziol1998} with a high star formation rate and NGC 6699 has a slightly higher star formation rate compared to the rest of the sample. There is some discrepancy between the radio continuum and the UV star formation rates for NGC 6808 and NGC 6699. These two galaxies have much higher calculated star formation rates based on the radio continuum than on the UV fluxes. A possible reason for this can be the different time scale of star formation rates that the UV and the radio continuum sample. UV emission originates from the hot young stars and samples recent star formation in the last $\sim$ 10-200 Myr \citep{Hao2011, Murphy2011}, whereas the 1.4 GHz continuum emission in star forming galaxies originates from super nova remnants tracing the star formation rates in the last $\sim$100 Myr \citep{Condon1992, Murphy2011}. Another possible explanation is additional radio emission from an active galactic nucleus (AGN) in NGC 6808 and NGC 6699. However this is unlikely, because when calculating the expected 1.4 GHz flux from the radio continuum - far-infrared (FIR) correlation we get similar estimates to the measured 1.4 GHz flux. We obtained the FIR luminosities from \cite{Li2011} and used q = 2.35 the average value for star forming late-type galaxies \citep{Yun2001}. We exclude the possibility of the higher 1.4 GHz flux originating from background radio galaxies, because we detect all catalogued radio galaxies in the field as separate sources to our sample galaxies (Fig.~\ref{fig:continuum}). 
	
Based on the UV star formation rates, we derived gas consumption times the following way: 
\begin{equation}
t_{gas} = 2.3 (M_{HI}/SFR_{UV,IR}),
\end{equation} 
where 2.3 is a correction factor accounting for the helium and the average molecular gas fraction of the ISM, assuming a uniform ratio between the combined molecular and neutral gas \citep{Young1996}. We find that all galaxies have short gas consumption times, comparable to starburst galaxies, between 0.1 - 0.6 Gyr. The average gas consumption time for local star forming galaxies is 1.5 Gyr \citep{Genzel2010}. The normal star formation rates combined with the low gas cycling times are suggestive of relatively recent and efficient gas removal. This excludes starvation as a dominant process responsible for the \HI-deficiency.

\begin{table*}
\caption{GALEX NUV and FUV, WISE 22$\mu$m and ATCA 1.4 GHz continuum fluxes for our sample and calculated star formation rates. SFR$_{UV,IR}$ is calculated following \citep{Hao2011} using FUV and 22$\mu$m flux and SFR$_{1.4GHz}$ is calculated following \citep{Murphy2011}. Gas depletion times are calculated based on SFR$_{UV,IR}$.}
\begin{tabular}{l c c c c c c c}
\hline
Galaxy name& GALEX NUV & GALEX FUV & WISE 22$\mu$m & ATCA 1.4 GHz & SFR$_{UV,IR}$ & SFR$_{1.4GHz}$ & gas depletion\\
& [ergs s$^{-1}$] & [ergs s$^{-1}$] & [ergs s$^{-1}$] & [mJy] & [M$_{\odot}$yr$^{-1}$] & [M$_{\odot}$yr$^{-1}$] & [10$^9$ year ]\\
\hline
IC 1993 & 1.63$\times 10^{42}$ & 1.45$\times 10^{42}$ & 7.13$\times 10^{41}$ & $<$ 3 & 0.2 $\pm$ 0.1 & $<$ 0.1& 0.2\\
NGC 1515 & 9.31$\times 10^{41}$ & 7.62$\times 10^{41}$& 1.06$\times 10^{42}$ & 21 $\pm$  2 & 0.2  $\pm$ 0.1& 0.3 & 0.6 \\
NGC 6808 & 4.51$\times 10^{42}$	& 3.47$\times 10^{42}$ & 2.11$\times 10^{43}$ & 64 $\pm$ 3 & 4.3 $\pm$ 0.5 & 10.3 & 0.1 \\
NGC 1473 & 1.90$\times 10^{42}$ & 2.24$\times 10^{42}$  & 4.29$\times 10^{41}$  & $<$ 3 & 0.2 $\pm$ 0.1& $<$ 0.1 & 0.4\\
NGC 6699 & 1.21$\times 10^{43}$ & 1.14$\times 10^{43}$ & 1.08$\times 10^{43}$ & 24 $\pm$  1 & 2.6  $\pm$ 0.4 & 4.0 & 0.1\\
ESO 009- G 010 & 3.02$\times 10^{42}$ &	- & 3.87$\times 10^{42}$ & 8 $\pm$  1 & 0.7 $\pm$ 0.1 & 0.6 & 0.5 \\
\hline
\end{tabular}
\label{tab:star-formation}
\end{table*}

\section{Description of the individual sample galaxies}

Our six sample galaxies are located in a variety of moderate density environments. One galaxy is in the outskirts of the Fornax cluster, four are in loose galaxy groups and one is in a triplet. We investigated the distance to the closest neighbouring galaxies in NED. None of the sample galaxies have significant neighbours within 200 kpc projected distances. We find that NGC 1515 and ESO 009-G 010 have dwarf galaxies within 200 kpc, but the rest of the sample does not have nearby neighbours. Their closest catalogued neighbours are located 400-500 kpc away. In the following sections we examine the individual sample galaxies in detail presenting information from the literature and the results from our \HI\ observations (see also Table~\ref{tab:signatures}). We also discuss the possible mechanisms responsible for the low \HI\ content of each galaxy based on their properties and environment.

\subsection{An \HI-deficient galaxy in the outskirts of a galaxy cluster}

IC 1993 is located in the outskirts of the Fornax cluster \citep{Fouque1992} at a projected distance of 850 kpc (2.5$^{\circ}$) from NGC 1399 the central galaxy of the cluster. Fornax is the second largest galaxy cluster in the local universe ($\lesssim$ 20 Mpc) after the Virgo cluster. Similar to Virgo, Fornax is an X-ray bright galaxy cluster \citep{Eckert2011} and has several \HI-deficient members \citep{Schroder2001, Waugh2002}. In contrast with Virgo, Fornax is a dynamically more evolved cluster and is about ten times less massive with $7\pm 2 \times 10^{13}$ \Msun\ \citep{Drinkwater2001}. The estimated mass of the Virgo cluster is $1.7 \times 10^{15}$ \Msun\ \citep{Fouque2001}. In terms of environment this makes Fornax more similar to large galaxy groups than to dense clusters \citep{Tonry2001}. The centre of the Fornax cluster is to the south-west of IC 1993. For the distance of IC 1993 we adopt the distance of the Fornax cluster 19.3 Mpc \citep{Tonry2001}. 

IC 1993 is a nearly face on galaxy with an inclination of 29$^{\circ}$ (HyperLEDA; \citealt{LEDA}) and a lopsided stellar disc \citep{Li2011}. It has an \HI\ mass of $2 \times 10^{8}$\Msun\ and an \HI-deficiency factor of 1.09 corresponding to $\sim$12 times less \HI\ than expected from its \textit{R}-band magnitude. The \HI\ morphology of this galaxy is strongly truncated. This galaxy has the smallest \HI\ to optical diameter ratio, with its \HI\ diameter being less than half of its optical diameter (Table~\ref{tab:diameters}). The \HI\ distribution is lopsided with the peak of the distribution located on the western side of the galaxy (Fig.~\ref{fig:HI-distribution}). The \HI\ velocity field of the galaxy shows signs of a rotating disk but is somewhat peculiar. The ATCA \HI\ flux is 70\% of the HIPASS \HI\ flux, which indicates some diffuse \HI\ emission around the galaxy. This is also supported by the HIPASS \HI\ line profile (Fig.~\ref{fig:HI-profiles}), with the excess emission on the receding side of the galaxy. However, IC 1993 is only a marginal detection in HIPASS, with a very noisy spectra, and the integrated \HI\ flux and \HI\ profile measured with the Nan\c{c}ay radio telescope (1.57 $\pm$ 0.53 Jy \kms, \citealt{Theureau2005}) are in good agreement with the ATCA data. Which makes the presence of diffuse gas questionable. There is also no significant difference between the integrated flux in the ATCA 750m and 1.5k observations. 

The lopsided \HI\ and stellar disk of this galaxy can indicate tidal interactions, similar to the fly-by scenario in the N-body simulations of \cite{Mapelli2008}. However, the strong \HI-deficiency and the truncated \HI\ disk of IC 1993 are more similar to face on galaxies experiencing ram pressure stripping in the Virgo cluster \citep{Chung2009}. In this case, a past passage through the central regions of the cluster, where a large fraction of the \HI\ disk was removed by ram pressure stripping, is a possible scenario for the gas removal. Ram pressure stripping is also supported by the presence of hot intra cluster medium in the centre of Fornax. Based on the \HI\ and stellar morphology, the strong \HI-deficiency and the location of this galaxy, the most likely gas removal mechanism is ram pressure stripping.

\begin{table*}
	\caption{Optical diameters are RC3 25 B mag arcsec$^{-2}$ diameters. \HI\ diameters are measured at 1\Msun pc$-1$. The last two columns show the ratios of the flux in the receding side and the approaching side of the \HI\ profile. Flux ratios $>$ 1.15 are considered slightly asymmetric and flux ratios $>$ 1.5 are considered strongly asymmetric. $ \langle I_{1}/I_{0}  \rangle _{B}$ is the strength of the lopsidedness parameter in the B-band \citep{Ho2011}.}
	\begin{tabular}{l c c c l l l}
		\hline
		Galaxy name & D$_{25}$ & D$_{HI}$ & D$_{HI}$/D$_{25}$  & flux ratio ATCA & flux ratio HIPASS & $ \langle I_{1}/I_{0}  \rangle _{B}$ \\
		& [arc min] & [arc min]  &   & & & \\
		\hline
		IC 1993 & 2.45 & 1.0 & 0.41 & 1.01 (sym) & 2.16 (strongly asym) & 0.47 $\pm$ 0.22\\
		NGC 1515 & 5.25 & 3.67 & 0.7 & 1.03 (sym) &  1.23 (asym) & 0.15 $\pm$ 0.01\\
		NGC 6808 & 1.51 & 1.0 & 0.66 & 1.59 (strongly asym) & 2.14 (strongly asym) & -\\
		NGC 1473 &  1.48 & 1.1 & 0.74 & 1.22 (asym) & 1.01 (asym) & - \\
		NGC 6699 & 1.55 & 1.2 & 0.77  & 1.24 (asym) & 1.24 (asym) & 0.27 $\pm$ 0.08\\
		ESO 009- G 010 & 2.09 & 2.0 & 0.95 & 1.06 (sym) & 1.23 (asym) &  0.20 $\pm$ 0.10\\
		\hline
	\end{tabular}
	\label{tab:diameters}
\end{table*}

\subsection{\HI-deficient galaxies in loose groups}

\subsubsection{NGC 1515}

NGC 1515 is a member of the Dorado group \citep{Fouque1992} and within the NGC 1566 subgroup \citep{Kilborn2005}. The Dorado group is one of the richest galaxy groups in the southern hemisphere, and is dominated by late-type galaxies. The group has X-ray emitting hot IGM \citep{Osmond2004}. \cite{Kilborn2005} previously identified two \HI-deficient galaxies in this group, NGC 1515 and NGC 1536, with 5-10 times less \HI\ than expected. The centre of the galaxy group is to the south-east from NGC 1515.

NGC 1515 is a nearly edge-on galaxy with an \HI\ mass of $5.2 \times 10^{8}$\Msun\ and an \HI-deficiency factor of 1.01 corresponding to $\sim$10 times less \HI\ than expected. The ATCA \HI\ flux is 99\% of the remeasured HIPASS \HI\ flux, which means that there is no significant amount of diffuse \HI\ gas around the galaxy. The \HI\ diameter of the galaxy is 0.7 times its optical diameter (Table~\ref{tab:diameters}) . The \HI\ velocity distribution shows a regularly rotating disk (Fig.~\ref{fig:velocity-distribution}) and the \HI\ profile of the galaxy is symmetric (Fig.~\ref{fig:HI-profiles}).         

The relatively symmetric \HI\ morphology, the regularly rotating \HI\ disk and the lack of diffuse \HI\ emission indicate that the gas removal did not occur recently, and the gas had time to settle back onto the galaxy. The symmetric stellar disk of this galaxy makes tidal interaction unlikely for the gas removal. \cite{Kilborn2005} suggest that NGC 1515 might be a `backsplash' galaxy, i.e. it passed through the centre of the galaxy group earlier where it lost a large fraction of its \HI\ and then moved towards the outskirts of the group, where it is located now. This scenario is supported by a faint \HI\ extension towards the south-east in the HIPASS data, the slightly asymmetric \HI\ profile and the presence of hot IGM in this group.

\subsubsection{NGC 6808}

NGC 6808 is a member of the Pavo galaxy group (NGC 6876 group, \citealt{deVaucouleurs1975}). The Pavo group is a moderately massive, dynamically young galaxy group with hot X-ray emitting intra cluster gas connecting its two brightest galaxies \citep{Machacek2005}. There are also several signs of recent and past interactions of galaxies in this group \citep{Machacek2009}. The centre of the galaxy group is to the south-east from NGC 6808. 

\begin{figure}
	\includegraphics[width=84mm]{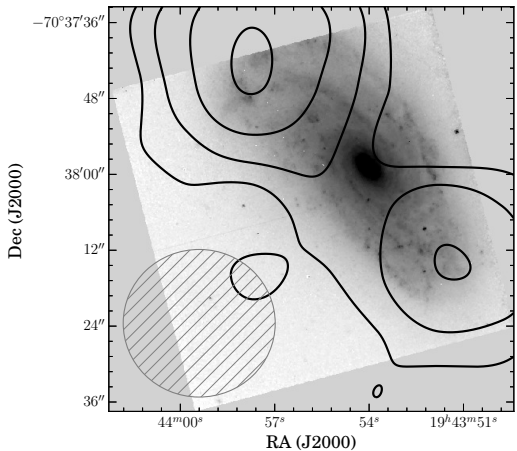}
	\caption{HST (NICMOS, F160W) image of NGC 6808 showing the warped stellar disk overlaid with the robust weighted \HI\ map. \HI\ contours are 16, 32, 64, 128  $\times 10^{19}$cm$^{-2}$.}
	\label{fig:NGC6808-Hubble}
\end{figure}

NGC 6808 is a highly inclined galaxy with an inclination of 50$^\circ$. It was identified as a starburst galaxy by \cite{Coziol1998}. The \HI\ mass of the galaxy is $2.5 \times 10^{9}$\Msun\ with an \HI-deficiency factor of 0.5, corresponding to $\sim$3 times less \HI\ than expected. The ATCA \HI\ flux is 78\% of the remeasured HIPASS \HI\ flux, which indicates diffuse \HI\ gas around the galaxy. This is also supported by the excess emission on the approaching side of the HIPASS \HI\ profile (Fig.~\ref{fig:HI-profiles}). The \HI\ distribution of this galaxy is strongly lopsided, with the peak of the distribution shifted towards the north-eastern side of the galaxy (Fig.~\ref{fig:HI-distribution}). This asymmetry is also visible in the velocity field and in the \HI\ profile, with more gas in the approaching side. NGC 6808 also appears to be asymmetric in the high-resolution HST NICMOS image (Fig.~\ref{fig:NGC6808-Hubble}). The north-eastern side of the galaxy is warped, indicating strong tidal interactions in the recent past of the galaxy.      

Considering the stellar warp, the \HI\ morphology and kinematics of NGC 6808 the most likely cause of \HI-deficiency is a  recent tidal interaction with another member of the galaxy group. Further support for this, is the starburst in the galaxy. There are several galaxies within 2$^{\circ}$ ($\sim$1.5 Mpc) and $\pm$ 800 \kms\ which are likely candidates for the interaction. The two closest similar sized galaxies are IC 4892, an edge-on spiral galaxy, and IC 4899 a slightly disturbed spiral galaxy. Tidal interactions can cause gas deficiency, distort the stellar and gaseous disk of galaxies and induce star formation (e.g. \citealt{Larson1978, Kennicutt1987}). Ram pressure stripping is not expected to cause a warp in the stellar disk of a galaxy. 

\subsubsection{NGC 1473}

NGC 1473 is classified as an irregular galaxy and as a member of a small galaxy group \citep{Garcia1993}. Beside NGC 1473 there are three spiral galaxies in the group, NGC 1511, NGC 1511A and NGC 1511B, located towards the north-east at a projected separation of $\sim$400 kpc. Observations of \cite{Koribalski2005} show that NGC 1511 has a disturbed \HI\ morphology and is currently interacting with NGC 1511B. NGC 1511 also has a peculiar stellar feature which is probably a remainder of a previously consumed galaxy. NGC 1511A does not show any signs of interaction. 

\begin{figure}
	\includegraphics[width=84mm]{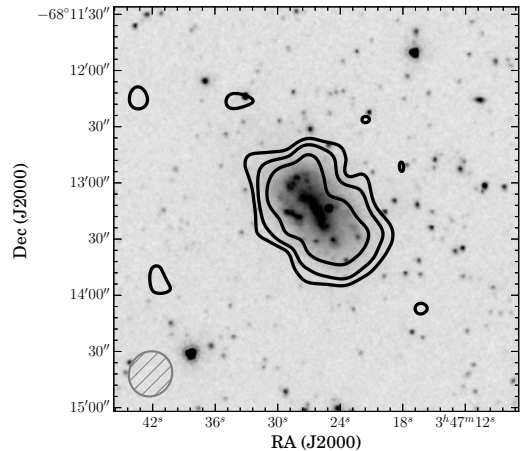}
	\caption{Spitzer S4G image of NGC 1473 overlaid with the the robust weighted \HI\ map. \HI\ contours are 16, 32, 64 $\times 10^{19}$cm$^{-2}$. The stellar bar in the centre is clearly visible. A large star forming clump is also visible on the south-eastern side of the galaxy. This star forming region is especially pronounced in the GALEX UV images of the galaxy.}
	\label{fig:NGC1473-S4G}
\end{figure}

NGC 1473 has an \HI\ mass of $3.8 \times 10^{8}$\Msun\ and an \HI-deficiency factor of 0.99 corresponding to $\sim$10 times less \HI\ than expected. The ATCA \HI\ flux is 75\% of the remeasured HIPASS \HI\ flux, i.e. about a fourth of the \HI\ gas is diffuse. The ratio of the integrated flux of the ATCA 750m and 1.5k observation is 0.64. This can imply recent gas removal, after which the gas did not have time to fully settle back onto the galaxy. The \HI\ diameter of NGC 1473 is 0.7 times its optical diameter and the \HI\ profile is asymmetric. The distribution of the \HI\ in the high resolution robust weighted data is slightly shifted towards the south-east from the centre of the galaxy (Figs.~\ref{fig:NGC1473-S4G} and \ref{fig:HI-distribution_highres}). The \HI\ velocity distribution shows a rotating disk. In the high resolution infrared S$^4$G image a central bar is visible (Fig.~\ref{fig:NGC1473-S4G}), which might indicate tidal interactions. Bars in late-type galaxies can be induced by tidal interactions which then can remain stable in the galaxy (e.g. \citealt{Noguchi1987, Gerin1990, Skibba2012, Mendez-Abreu2012}). Beside the stellar bar, there is also a large star forming clump on the south-eastern side of the galaxy. This star forming region is especially pronounced in the GALEX UV images.

Possible gas removal scenarios for NGC 1473 are tidal interactions with other group members - the stellar bar could be a result of such interactions - or ram pressure stripping by the intragroup medium, which can be supported by the slightly off centre \HI\ distribution (Fig.~\ref{fig:NGC1473-S4G}).       

\subsubsection{NGC 6699}

NGC 6699 is an Sbc galaxy with an inclination of 27$^{\circ}$ and a lopsided stellar disk \citep{Li2011}. It is a member of a moderate sized galaxy group, with 27 members \citep{Giuricin2000, Crook2007}. \cite{Arp1987} and \cite{Casasola2004} classify this galaxy as part of a pair of interacting galaxies. The nearest neighbour is IC 4775 an \HI\ deficient ($DEF = 0.45$) galaxy at a projected separation of $\sim$ 400 kpc to the north-west. The centre of the galaxy group is to the north-east from NGC 6699.

NGC 6699 has an \HI\ mass of $1.98 \times 10^{9}$\Msun\ and an \HI-deficiency factor of 0.76 corresponding to $\sim$6 times less \HI\ than expected. The ATCA \HI\ flux is 66\% of the remeasured HIPASS \HI\ flux, which means that a significant fraction of the \HI\ gas is diffuse. The flux ratio between the ATCA 750m and 1.5k measurement is 0.63. The diffuse \HI\ can be indicative of recent gas removal. The \HI\ diameter is smaller than the optical diameter and the \HI\ morphology is slightly lopsided, with the peak of the distribution towards the southern side of the galaxy (Fig.~\ref{fig:HI-distribution}). The HIPASS data also shows a faint extension towards the south. The \HI\ profile is asymmetric. The velocity field shows a regularly rotating \HI\ disk.

The lopsided stellar and \HI\ disk, the large amount of diffuse \HI\ and the relatively enhanced star formation compared to the other sample galaxies are indicative of tidal interactions. The stellar disk of this galaxy looks very similar to the `fly-by' case of the simulation by \cite{Mapelli2008}. In addition, the closest nearby galaxy IC 4775 also appears to be \HI-deficient, making it a good candidate for past tidal interactions. Based on the \HI\ morphology, ram pressure stripping is also a possibility for gas removal. However, ram pressure stripping can not explain the enhanced star formation rate and the lopsided stellar disk.

\subsection{An \HI-deficient galaxy in a triplet}

ESO 009-G 010 has an inclination of 45$^{\circ}$ and is a member of a galaxy triplet \citep{Karachentseva2000}. The other two galaxies in the triplet are NGC 6438 an S0 galaxy and NGC 6438A an irregular galaxy. NGC 6438 and NGC 6438A are interacting with each other and are likely in the process of merging. ES0 009-G 010 is located $\sim 1^{\circ}$ to the north-west from the merging pair. This distance corresponds to $\sim 500$ kpc, which makes this a rather loose triplet.

ESO 009-G 010 has an \HI\ mass of $1.8 \times 10^{9}$\Msun\ and an \HI-deficiency factor of 0.76 corresponding to $\sim$6 times less \HI\ than expected. The ATCA \HI\ flux is 42\% of the remeasured HIPASS \HI\ flux, which indicates that about half of the \HI\ gas is diffuse. The HIPASS data of ESO 009-G 010 shows an extension towards the east and an asymmetric \HI\ profile. The \HI\ diameter is approximately the same size as the stellar diameter of the galaxy (Table~\ref{tab:diameters}). The \HI\ velocity field is somewhat peculiar, but shows a rotating disk. 

ESO 009-G 010 has a lopsided stellar disk \citep{Li2011}, which could be caused by gravitational interactions. The presence of diffuse \HI\ and a faint \HI\ extension in HIPASS towards the east is also indicative of tidal interactions. The \HI\ morphology could also be explained by ram pressure stripping. However, ram pressure stripping can not account for the lopsided stellar disk, and it is unlikely that a galaxy triplet has dense enough IGM for strong ram pressure stripping. 

\subsubsection{A newly discovered dwarf galaxy: ATCA1753-8517}

\begin{figure*}
	\includegraphics[width=17cm]{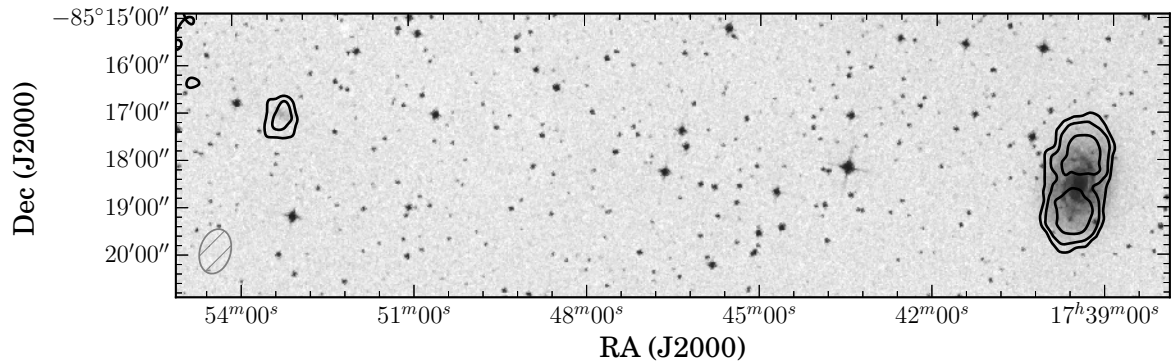}
	\caption{DSS2 blue image of ESO 009-G 010 and ATCA1753-8517 overlaid with \HI\ contours at 4, 8, 16 $\times 10^{19}$cm$^{-2}$.}
	\label{fig:Smudgy}
\end{figure*}

We report the discovery of a small, previously uncatalogued dwarf galaxy, ATCA1753-8517, in the ATCA data cubes of ESO 009-G 010. ATCA1753-8517 is located $\sim$200 kpc towards the east from ESO 009-G 010 and has a faint optical counterpart on the DSS 2 Blue image (Fig.~\ref{fig:Smudgy}). ATCA1753-8517 also has a counterpart  in the GALEX NUV image. We measure an \HI\ mass of 3.43$\times 10^{7}$ M$_{\odot}$.  Measured properties of ATCA1753-8517 are given in Table~\ref{tab:Smudgy}. It is a possibility that this galaxy could be a tidal dwarf galaxy, given that it lies in the centre of a galaxy triplet and that ESO 009-G 010 has six times less \HI\ than expected. Tidal dwarf galaxies are often found in major mergers but can also form in mildly interacting systems such as NGC 1512/1510 \citep{Koribalski2009}. Tidal dwarfs do not have their own dark matter halo, which means that their dynamic-to-\HI\ mass ratio is low compared to normal dwarf galaxies. To investigate this possibility we estimated the dynamical mass of ATCA1753-8517 using the following equation:
\begin{equation}
$$M_{dyn} = 3.39 \times a_{HI}d \left( \frac{w_{20}}{2} \right)^{2} M_{\odot}$$
\end{equation}
where $a_{HI}$ is the diameter of the galaxy in arc minutes, $d$ is the distance to it in Mpc and $w_{20}$ is the width at 20 \% of the peak flux density \citep{Giovanelli1988}. Since ATCA1753-8517 is unresolved in the \HI\ observation we assume a diameter of 50 arc seconds, corresponding to the size of the synthesised beam, which gives a dynamical mass of $1.1 \times 10^{9}$ \Msun. Assuming a gas mass of 1.3 times the \HI\ mass we get a $M_{dyn}$/$M_{gas} \sim 25$. Depending on the assumed diameter of the galaxy this can vary between 15-25. For local group dwarf irregular galaxies a $M_{dyn}$/$M_{gas}$ ratio of 10-20 is typical. This suggests that ATCA1753-8517 is a dwarf irregular galaxy with a dark matter halo and is not a tidal dwarf galaxy.   

\begin{table}
	\caption{Properties of the new dwarf galaxy ATCA1753-8517. }
	\begin{tabular}{l l }
		\hline
		Coordinates & 17:53:4.9, -85:17:31.9 \\
		\HI\ flux & 0.119 Jy \kms \\
		Velocity & 2507 \kms \\
		$w_{50}$ & 46 \kms \\
		$w_{20}$ & 65 \kms \\
		Distance & 35 Mpc \\
		Distance to ESO 009-G 010 & 188 kpc \\
		HI mass & $3.4 \times 10^{7} $ \Msun\\
		\hline
	\end{tabular}
	\label{tab:Smudgy}
\end{table}

\section{Discussion}
\label{Discussion}

In this section we discuss our results and compare them to high resolution observations of \HI-deficient galaxies from the literature.

The VLA Imaging of Virgo spirals in Atomic gas (VIVA, \citealt{Chung2009}) survey mapped 53 late-type galaxies in the Virgo cluster to study the effect of the cluster environment on the \HI\ content of galaxies. The typical resolution of the VIVA survey is 15'' with a column density sensitivity of $\sim 3-5 \times 10^{19}$ cm$^{-2}$. \cite{Chung2009} found that galaxies in the Virgo cluster exhibit a variety of \HI\ morphologies and properties. They found galaxies with truncated \HI\ disks, one sided \HI\ tails, galaxies with a lopsided \HI\ content and gas rich galaxies with extended \HI\ disks. A large fraction of the galaxies in Virgo exhibit signs of recent gas removal by ram pressure stripping or by tidal interactions. 

\begin{figure}
	\includegraphics[width=84mm]{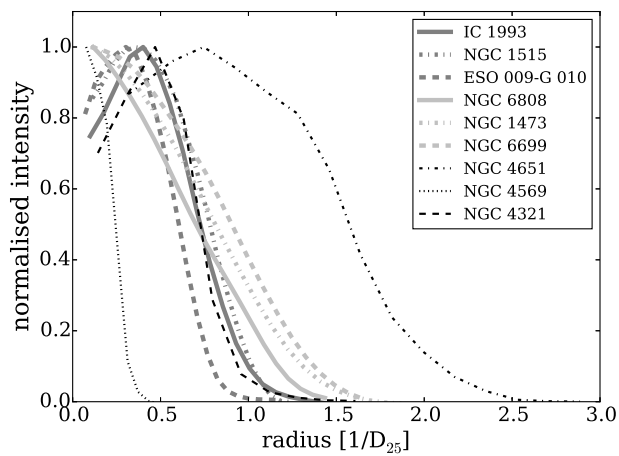}
	\caption{\HI\ surface brightness profiles of the sample galaxies compared to \HI\ surface brightness profiles of three Virgo cluster galaxies (NGC 4651, NGC 4569, NGC 4321).}
	\label{fig:surface-brigthness}
\end{figure}

Comparing our galaxies to the \HI-deficient galaxies in the VIVA survey we find similar signatures of \HI\ stripping - truncated \HI\ disks and lopsided gas distribution - but to a lesser extent. We do not observe \HI\ tails or tidal arms down to a column density limit of 4 $\times 10^{19}$ cm$^{-2}$. This column density sensitivity should be sufficient to detect tidal arms. However, due to our limited spatial resolution, which is $\sim$4 times less than VIVA's, we are not able to resolve small \HI\ tails.

To further compare our data with VIVA galaxies, we measure the average \HI\ radial surface brightness profiles of our sample using elliptic annuli centred on the optical image of the galaxy. In Fig.~\ref{fig:surface-brigthness} we compare the measured radial profiles with three galaxies (NGC 4569, NGC 4321 and NGC 4651) from the Virgo cluster \citep{Chung2009}. The comparison galaxies are selected to have different \HI\ properties, ranging from extremely \HI-deficient to a gas rich. NGC 4569 has a strongly truncated \HI\ disk, NGC 4321 has a similar sized \HI\ disk than its stellar disk and NGC 4651 is a gas rich galaxy with an extended \HI\ disk. The \HI\ profiles of our sample galaxies are most similar to NGC 4321. \cite{Cayatte1994} classified the \HI\ profiles of Virgo cluster galaxies into four groups based on their \HI\ to optical diameter ratio: galaxies with
	\begin{inparaenum} 
		\item  normal \HI\ distribution ($D_{HI}/D_{opt} > 1.3$),
		\item  slightly truncated \HI\ disks ($0.75 < D_{HI}/D_{opt} < 1.3$),
		\item  severely truncated \HI\ disks ($D_{HI}/D_{opt} < 0.75$),
		\item  low \HI\ surface densities with a central surface density hole.
	\end{inparaenum} 
In this classification, our sample galaxies fall into group II and III. \cite{Cayatte1994} conclude that the most likely gas removal mechanisms for these galaxies are ISM interactions such as ram-pressure stripping and viscous stripping. However, their results take into account the presence of the dense intra cluster medium in the Virgo cluster, which makes strong ram pressure stripping possible for most cluster galaxies. In contrast to the Virgo cluster, our sample galaxies are mainly located in relatively low density environments.

Other examples of resolved imaging of \HI\ deficient galaxies in clusters are VLA observations of spiral galaxies in the Coma cluster \citep{Bravo-Alfaro2001} and in the Pegasus I cluster \citep{Levy2007}. \HI\ deficient galaxies in these clusters have truncated \HI\ disks and lopsided \HI\ morphologies, which are interpreted as the result of ram pressure stripping and tidal interactions. Our sample galaxies show similar but less extreme signs of gas removal than the \HI\ deficient galaxies in Coma. 

In terms of \HI\ resolution and environment density a similar sample to our galaxies is the ATCA observations of 16 southern Group Evolution Muliwavelength Study (GEMS) sources \citep{Kern2008}. The 16 galaxies from this study are members of 6 different galaxy groups and many of them show signs of gas removal. For example warps, truncated \HI\ disks and extended \HI\ features. We calculated the \HI-deficiency parameter for the galaxies that have available optical data from HyperLEDA (10 galaxies) and found that based on their ATCA \HI\ measurement 50 \% are \HI-deficient ($DEF > 0.6$). Considering that ATCA might miss some of the \HI\ flux, we also calculated the deficiency parameter based on Parkes data from \cite{Kilborn2009}, where we find that 30 \% are \HI-deficient. These three galaxies are NGC 7204 - a strongly interacting close pair of galaxies -, IC 1953 and NGC 1385. IC 1953 and NGC 1385 have similar \HI\ morphologies to our sample galaxies. IC 1953 has a truncated \HI\ disk and NGC 1385 has a relatively small, warped \HI\ disk. Both galaxies have asymmetric stellar disks suggestive of past tidal interactions. 

\section{Summary and Conclusions}
\label{Summary}

We present new high-resolution ATCA \HI\ observations of six \HI-deficient galaxies identified from HIPASS using the \HI-optical scaling relations from \cite{paper1}. All sample galaxies are located in intermediate density environments. One galaxy is in the outskirts of a galaxy cluster, four are in loose galaxy groups and one is in a triplet.
 
We present the \HI\ properties of this galaxy sample including \HI\ distribution, velocity fields and \HI\ line profiles. Analysing multi-wavelength data - UV, optical, infrared and \HI\ imaging - we investigate the possible causes of the \HI-deficiency. All of the sample galaxies have smaller \HI\ diameters, measured at 1 \Msun pc$^{-2}$, than their B 25 mag arcsec$^{-2}$ optical diameter. In contrast to this, typical late-type galaxies have \HI\ diameters that are 1.7-2 times more extended than their optical diameter \citep{Broeils1994}. Two galaxies, IC 1993 and NGC 6808 show strong asymmetries and almost all galaxies have a lopsided \HI\ distributions. NGC 6808 has a warped stellar disk which in combination with the strongly asymmetric \HI\ disk, clearly indicate strong tidal interactions in the recent past of this galaxy. Half the sample galaxies have slightly lopsided stellar disks, which can suggest tidal interactions. In the case of two galaxies, ram pressure stripping is the most likely candidate for the gas removal and for two more galaxies it is a possibility, considering the \HI\ morphology and the environment of these galaxies. From this we conclude that both ram pressure stripping and tidal interactions are important gas removal mechanisms in low density environments. Based on the star formation rates and the gas cycling times of the sample, strangulation is unlikely to be the main reason for \HI-deficiency. 

Comparing our sample galaxies to other \HI\ observations, we find that they show similar \HI\ characteristics to \HI\ deficient galaxies in the Virgo cluster and in the GEMS groups. Similar to \cite{Kern2008}, we find that based on the \HI\ and optical morphology of our sample, the most likely cause of \HI-deficiency is gravitational interactions for four sample galaxies. Deeper optical observations and sensitive high resolution \HI\ imaging might clarify the cause of \HI-deficiency of these galaxies. 

From our observations, we conclude that gravitational interactions play an important role in shaping the \HI\ content of the sample galaxies. However, we also find that based on the \HI\ and optical morphology, ram pressure stripping may play an important part in pre-processing galaxies in low density environments. To investigate the relative importance of tidal interactions and ram pressure stripping in galaxy groups, a larger statistical sample is needed. Upcoming large scale \HI\ and optical surveys, such as the ASKAP \HI\ All Sky Survey, known as WALLABY (\citealt{Wallaby}; \citealt{Koribalski2012}) and SkyMapper \citep{SkyMapper2007} will provide us with large samples of resolved \HI\ maps of galaxies to answer this question.  

\section{Acknowledgements}

We would like to thank the anonymous referee for the useful comments and suggestions, which helped us to greatly improve the paper.

HD is supported by a Swinburne University SUPRA postgraduate scholarship.

OIW acknowledges a Super Science Fellowship from the Australian Research Council.

We would like to thank Barbara Catinella and Luca Cortese for helpful discussions on this work.

The Australia Telescope Compact Array and the Parkes telescope are part of the Australia Telescope National Facility which is funded by the Commonwealth of Australia for operation as a National Facility managed by CSIRO. 

This research has made use of the NASA/IPAC Extragalactic Database (NED) which is operated by the Jet Propulsion Laboratory, California Institute of Technology, under contract with the National Aeronautics and Space Administration. 

We acknowledge the usage of the HyperLeda database (http://leda.univ-lyon1.fr).

Part of this research was based on observations made with the NASA/ESA Hubble Space Telescope, and obtained from the Hubble Legacy Archive, which is a collaboration between the Space Telescope Science Institute (STScI/NASA), the Space Telescope European Coordinating Facility (ST-ECF/ESA) and the Canadian Astronomy Data Centre (CADC/NRC/CSA).

\appendix{}
\section{Additional Figures and Table}

\begin{figure*}
	\subfigure{\includegraphics[width=5.5cm]{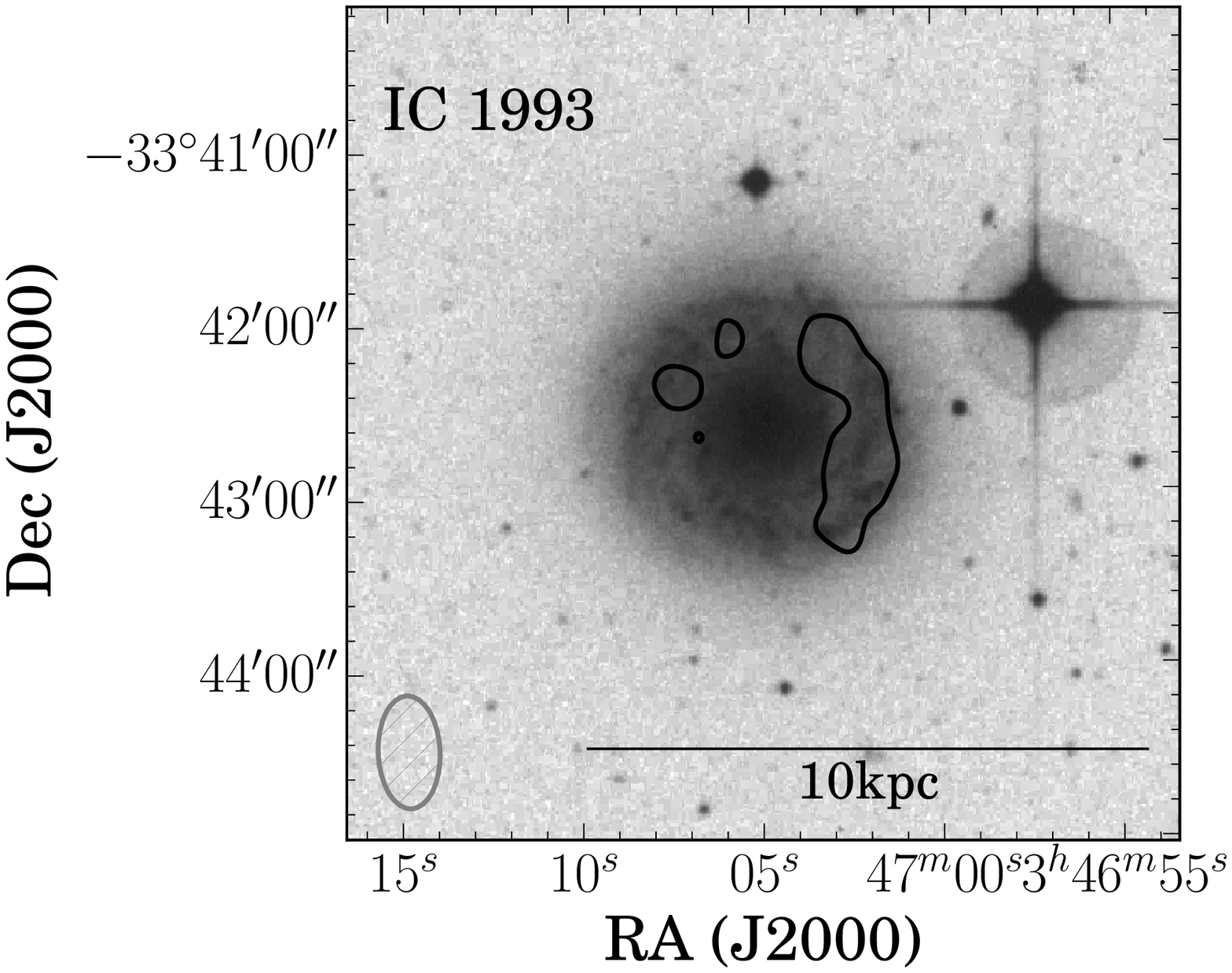}} 
	\subfigure{\includegraphics[width=5.5cm]{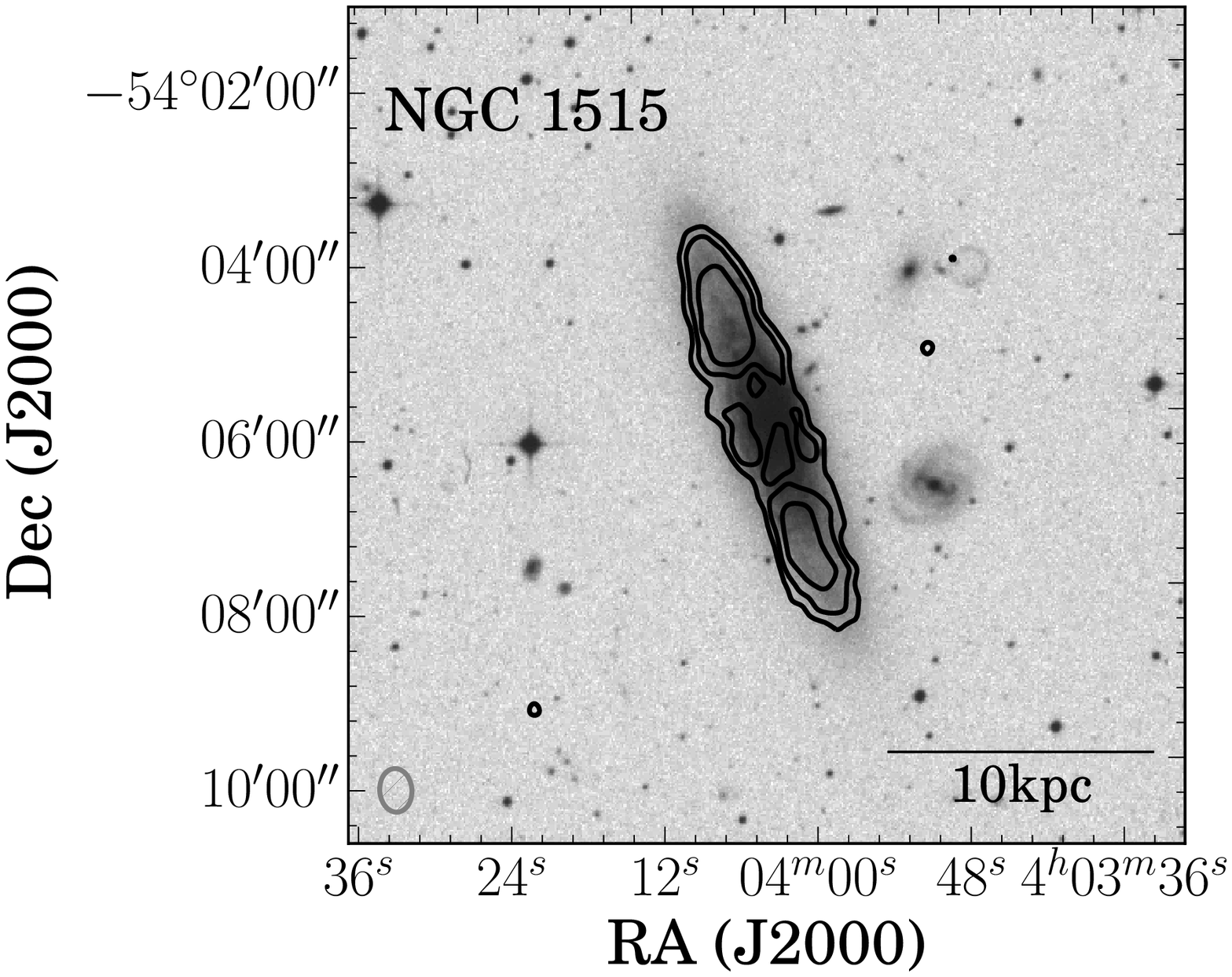}} 
	\subfigure{\includegraphics[width=5.5cm]{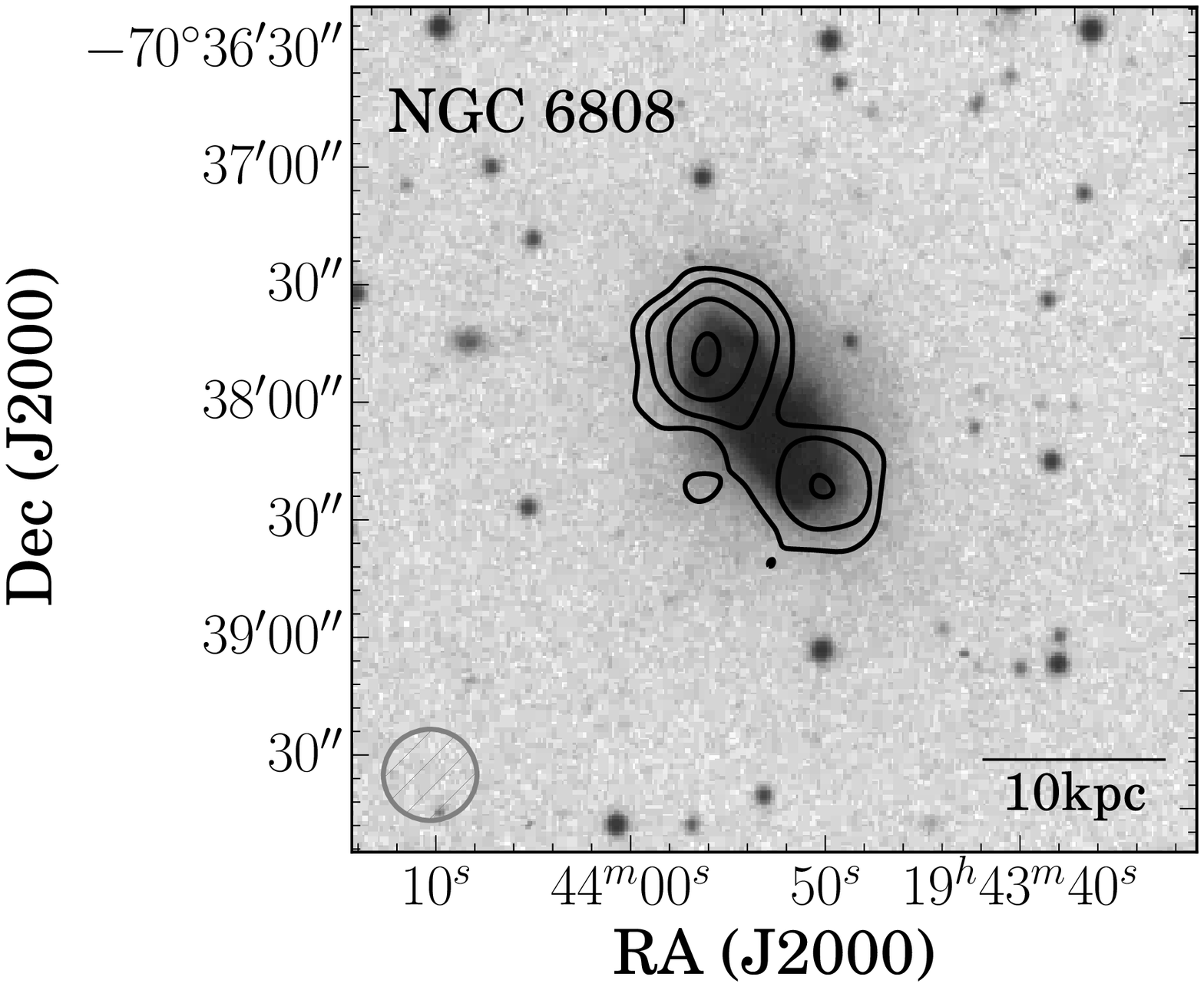}} 
	\subfigure{\includegraphics[width=5.5cm]{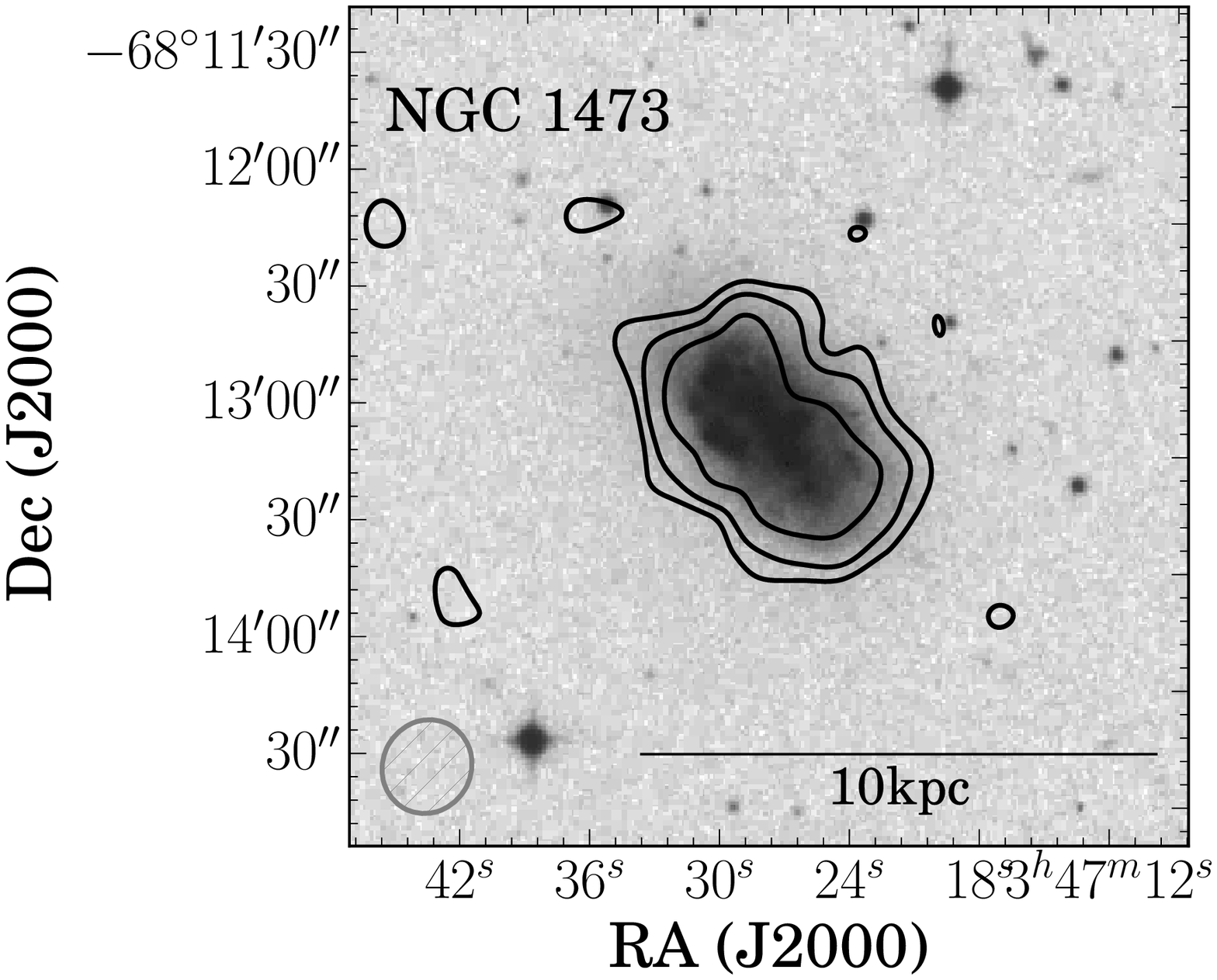}} 
	\subfigure{\includegraphics[width=5.5cm]{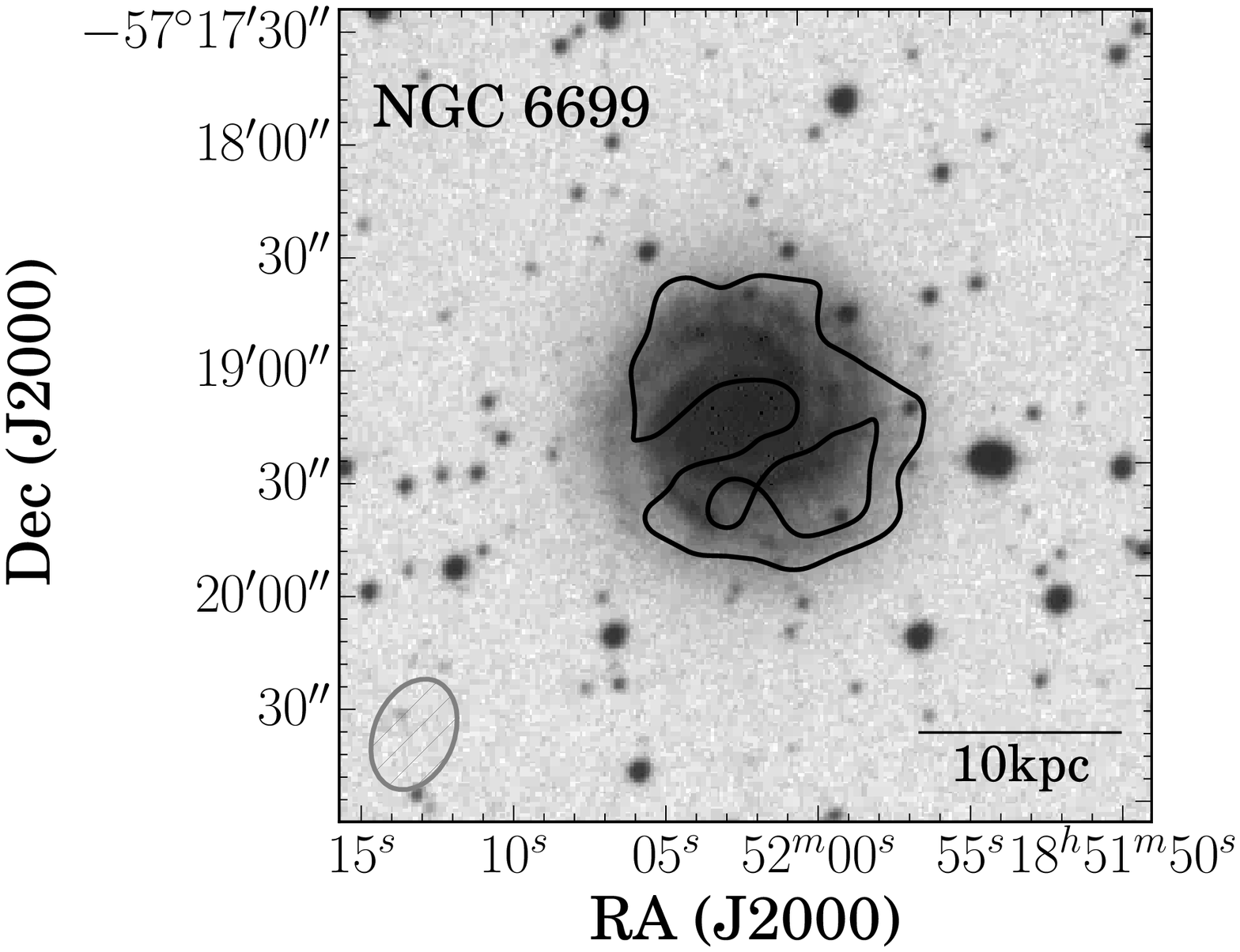}} 
	\subfigure{\includegraphics[width=5.5cm]{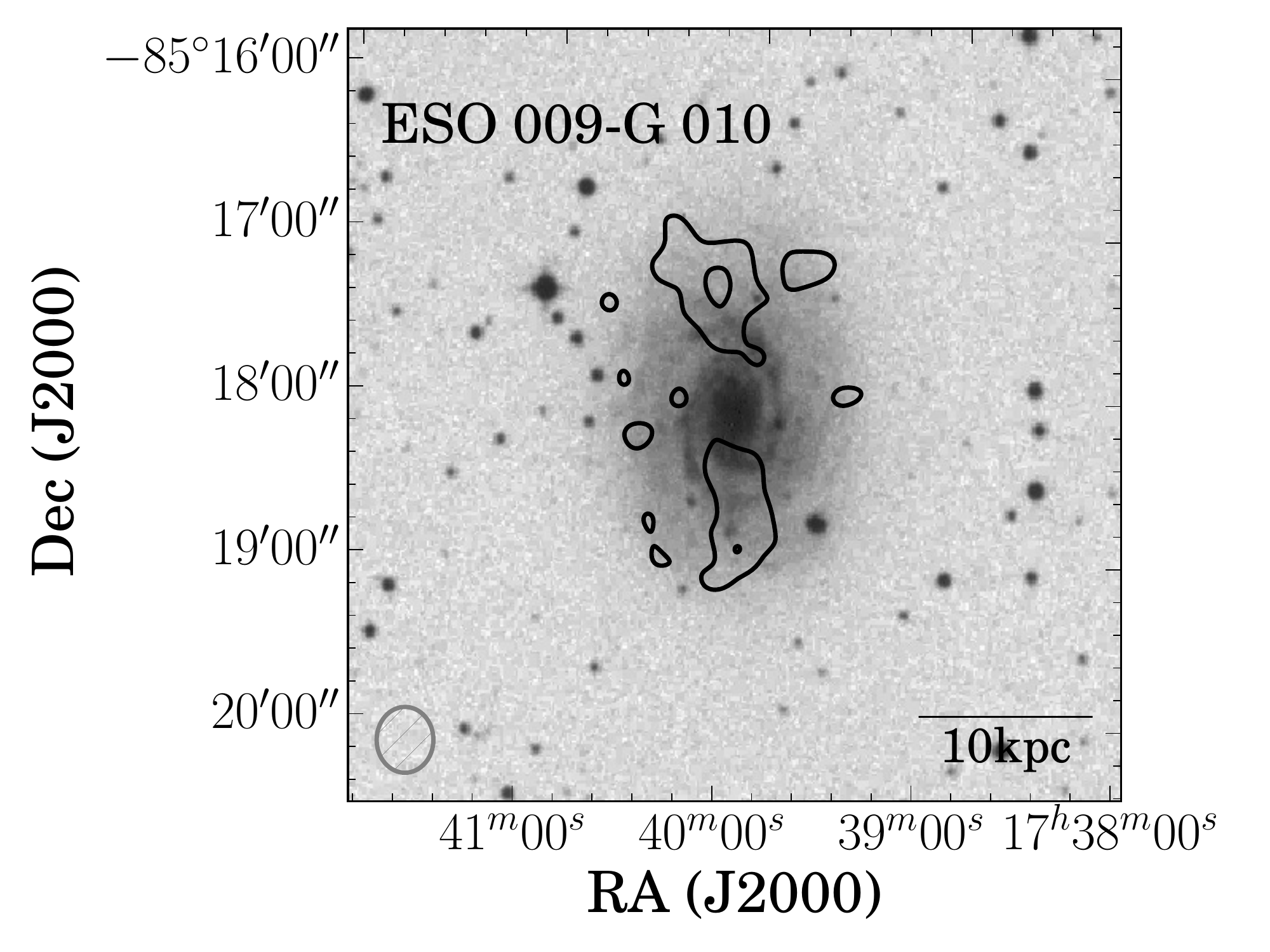}}  
	\caption{Same as Fig~\ref{fig:HI-distribution}, except higher angular resolution of the \HI\ data using robust weighting. Contour levels are 16, 32, 64, 128 $\times\ 10^{19}$cm$^{-2}$, except for IC 1993 where contour levels are 8 $\times\ 10^{19}$cm$^{-2}$.}
	\label{fig:HI-distribution_highres}
\end{figure*}

\begin{figure*}
	\subfigure{\includegraphics[width=5.5cm]{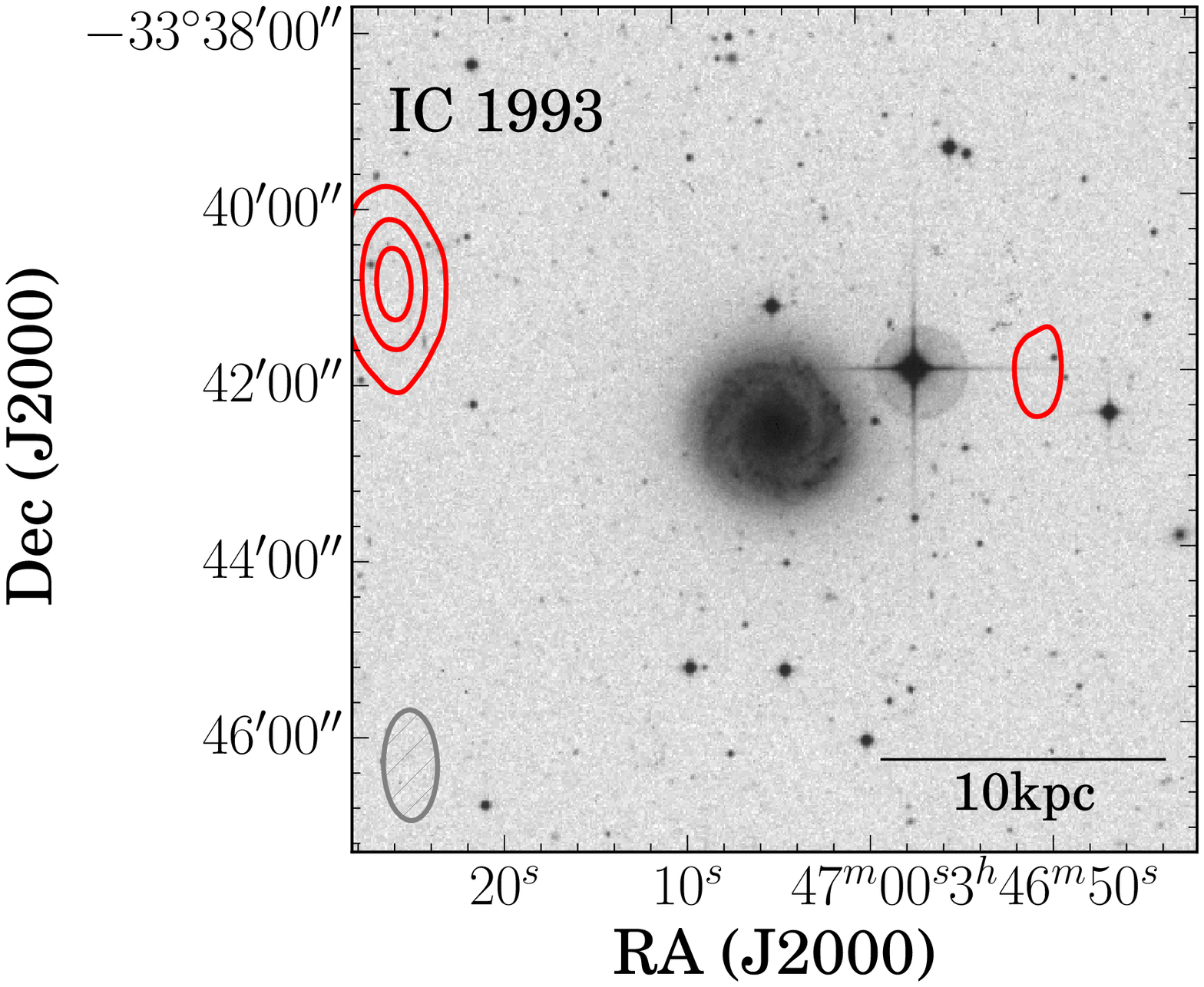}} 
	\subfigure{\includegraphics[width=5.5cm]{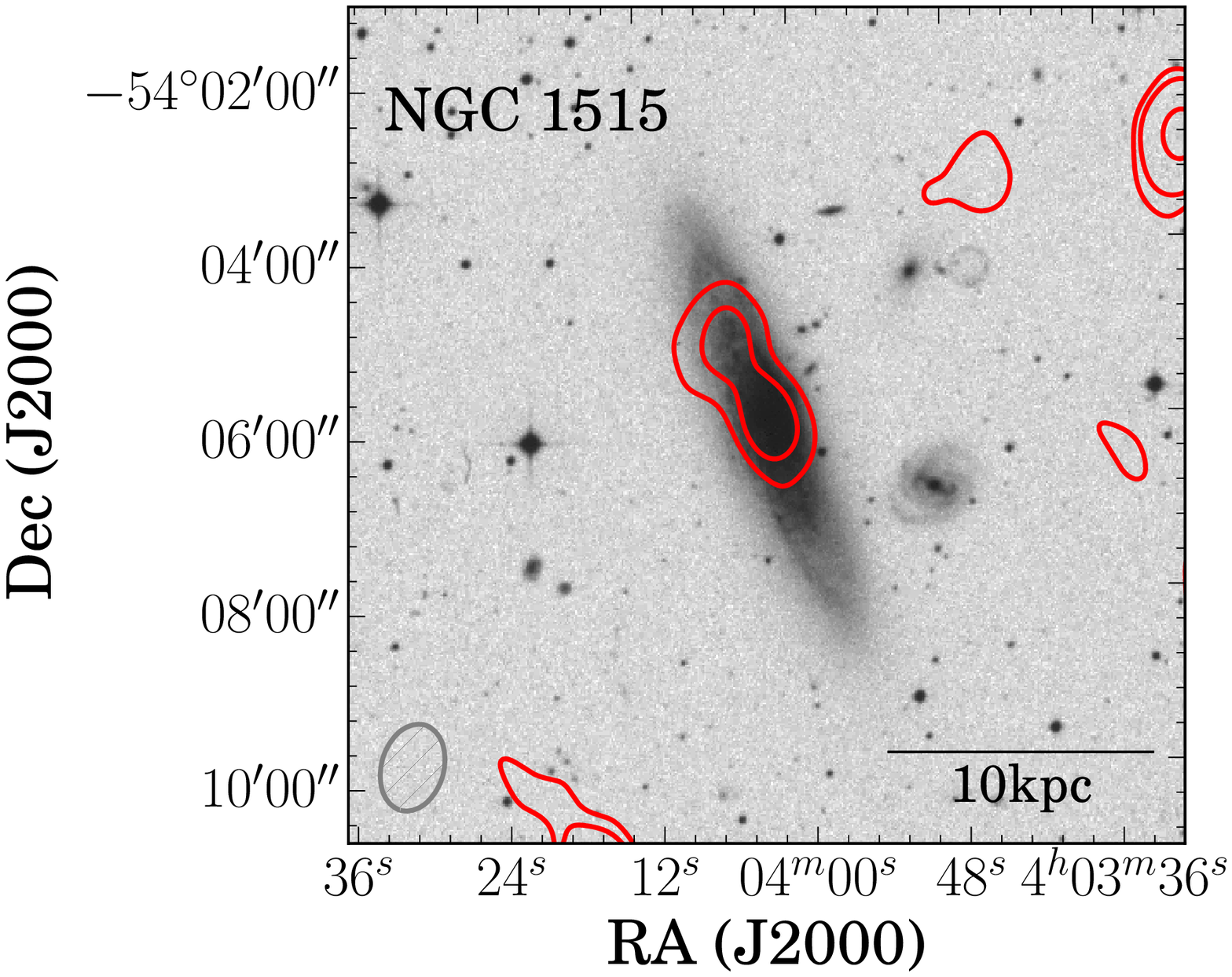}} 
	\subfigure{\includegraphics[width=5.5cm]{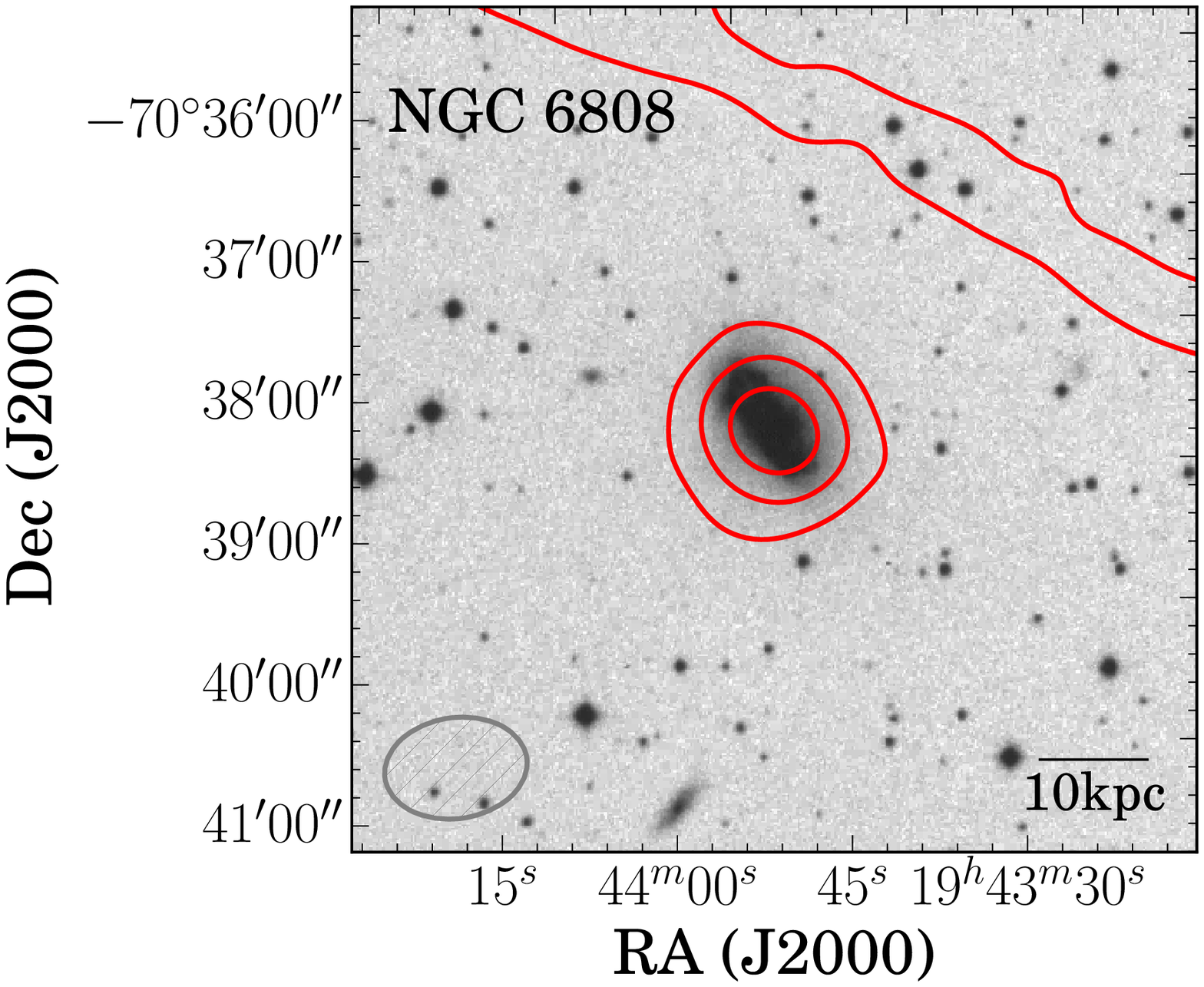}} 
	\subfigure{\includegraphics[width=5.5cm]{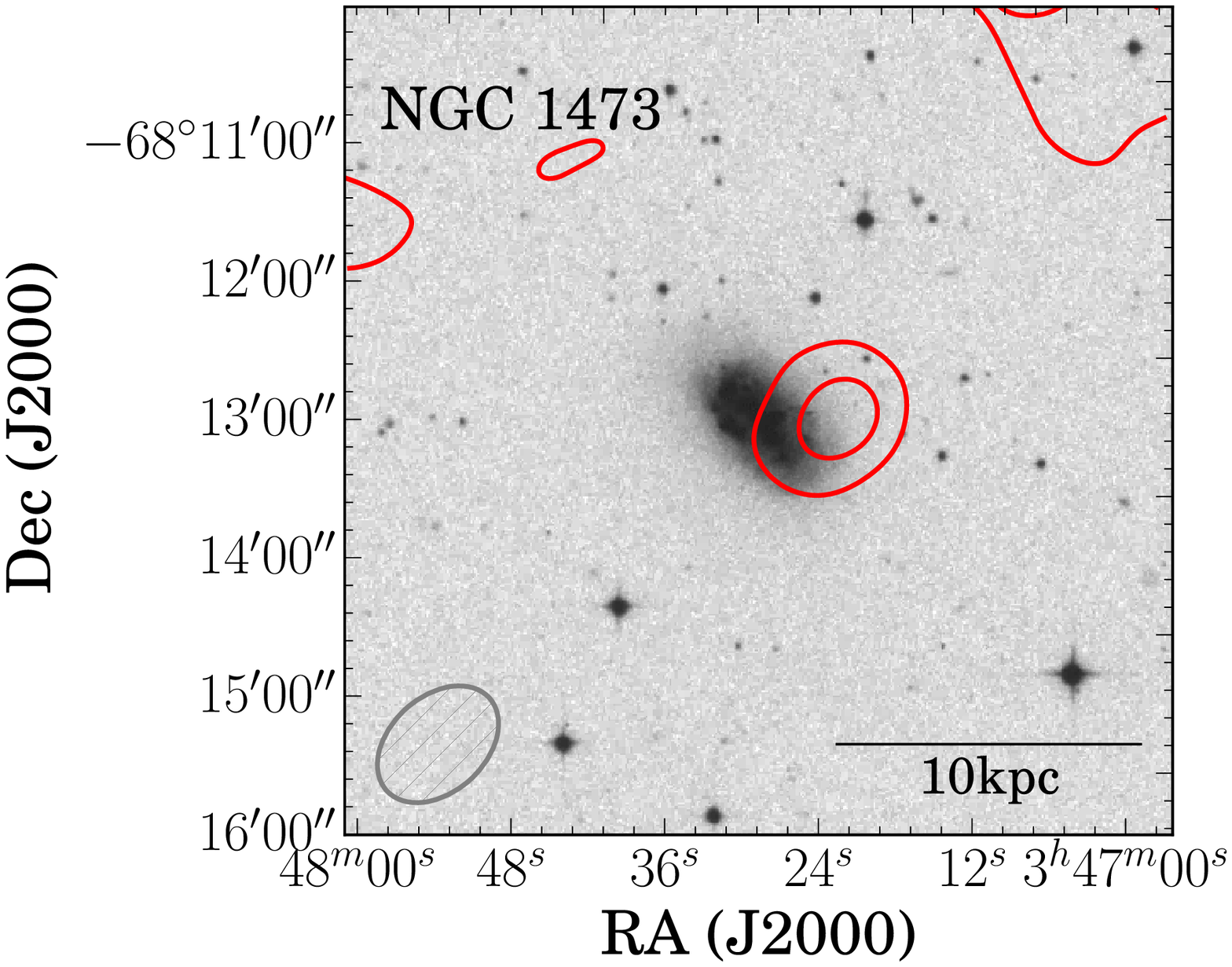}} 
	\subfigure{\includegraphics[width=5.5cm]{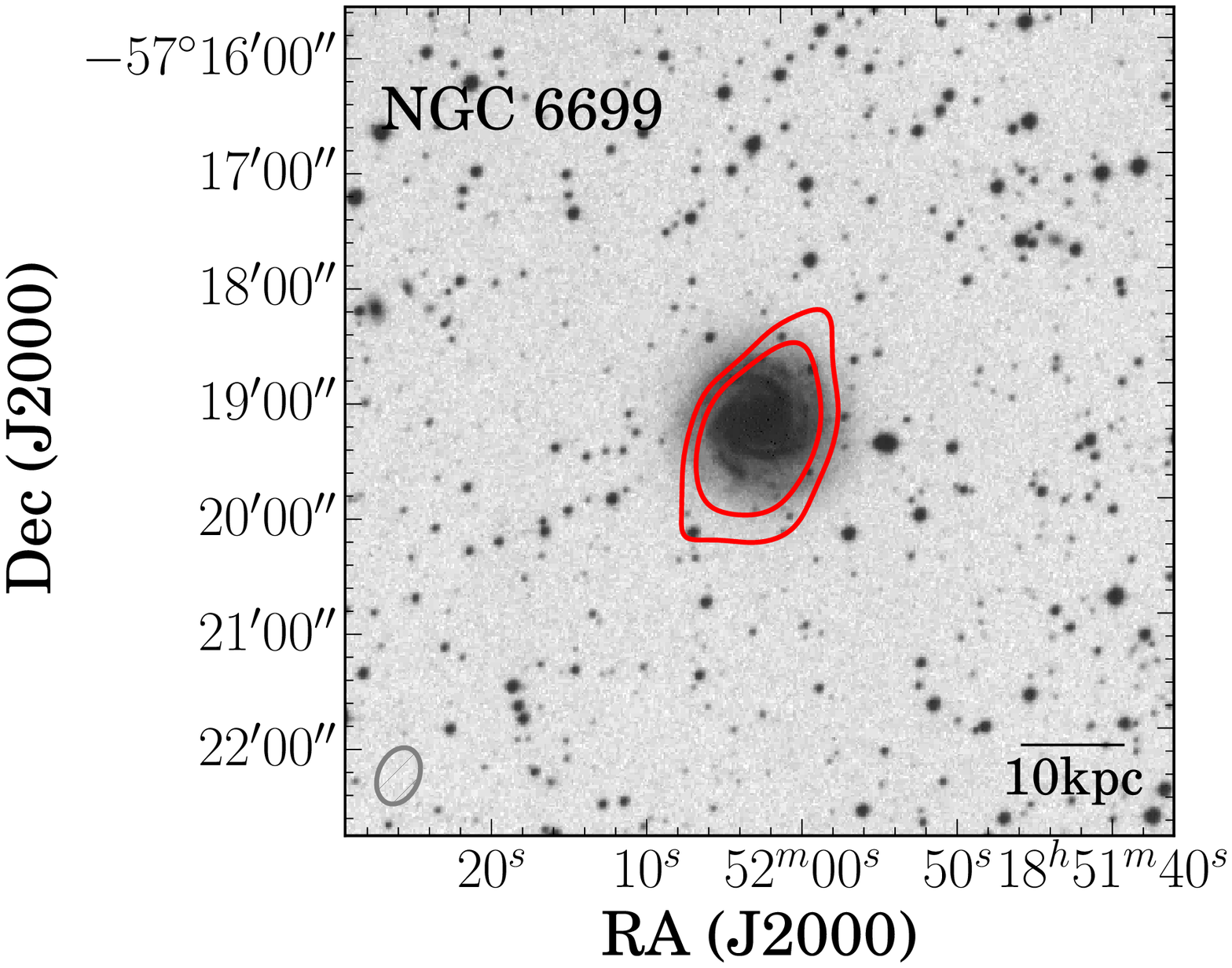}} 
	\subfigure{\includegraphics[width=5.5cm]{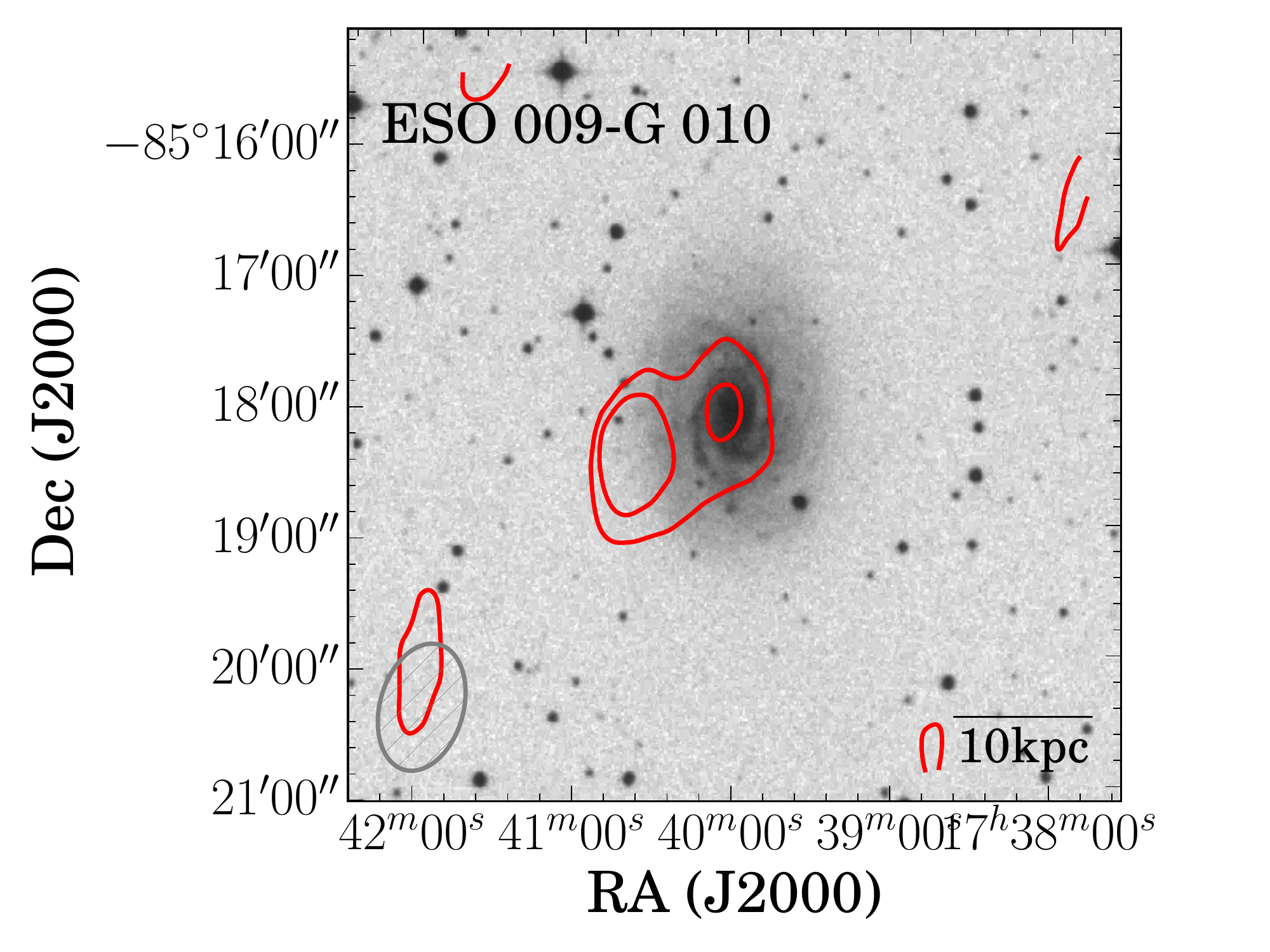}}  
	\caption{1.4 GHz continuum contours from the combined ATCA data overlaid onto optical B-band images from SuperCOSMOS (UKST). Contour levels are 0.004, 0.008, 0.016 Jy beam${-1}$ for IC 1993, NGC 6808 and NGC 1473; 0.002, 0.004, 0.008 Jy beam${-1}$ for NGC 1515, NGC 6699 and ESO 009-G 010. The synthesised beam size is in the lower left corner and size scale (10 kpc) is in the lower right corner of each image.}
	\label{fig:continuum}
\end{figure*}

\begin{figure*}
	\subfigure{\includegraphics[width=5.5cm]{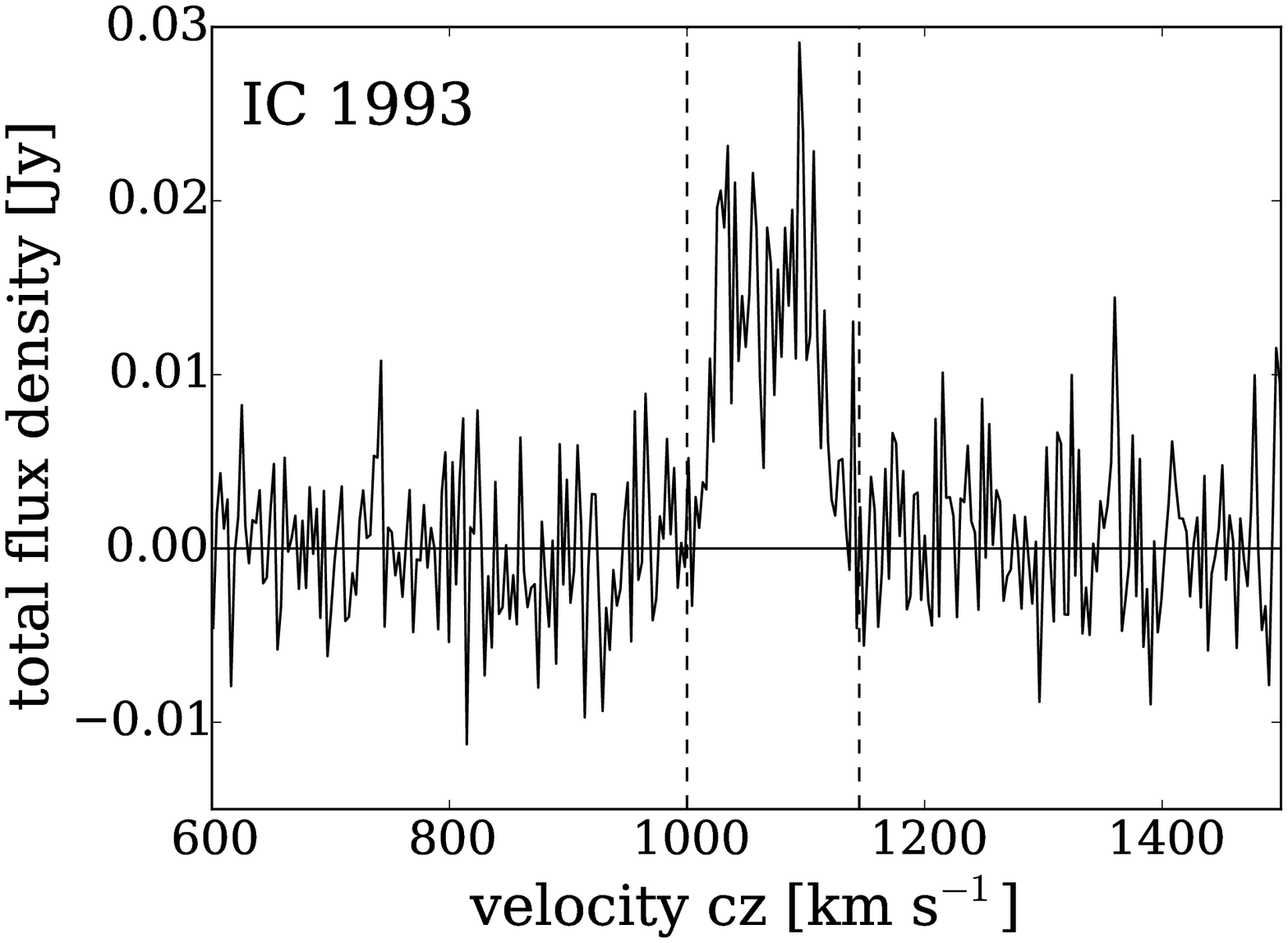}} 
	\subfigure{\includegraphics[width=5.5cm]{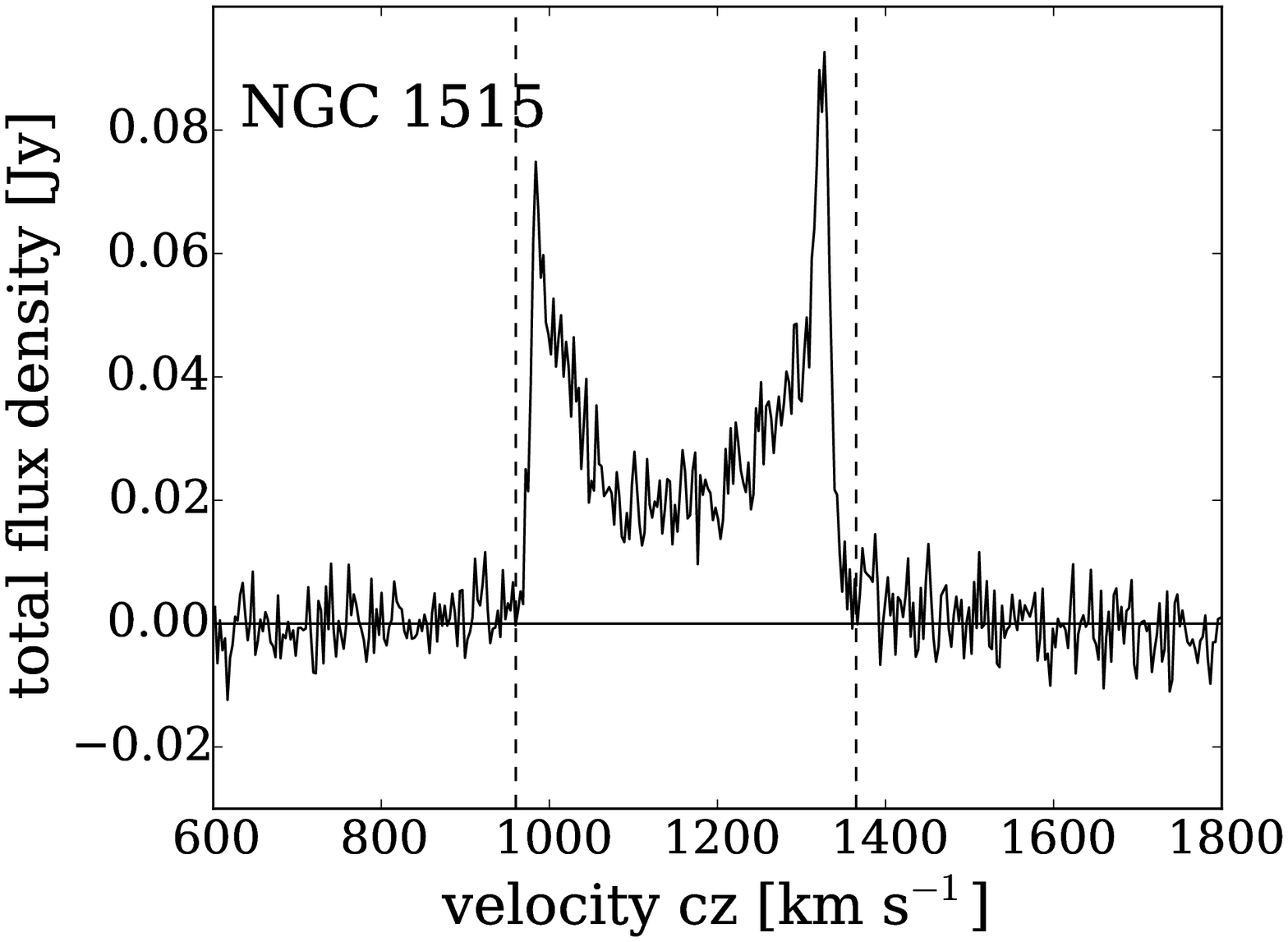}} 
	\subfigure{\includegraphics[width=5.5cm]{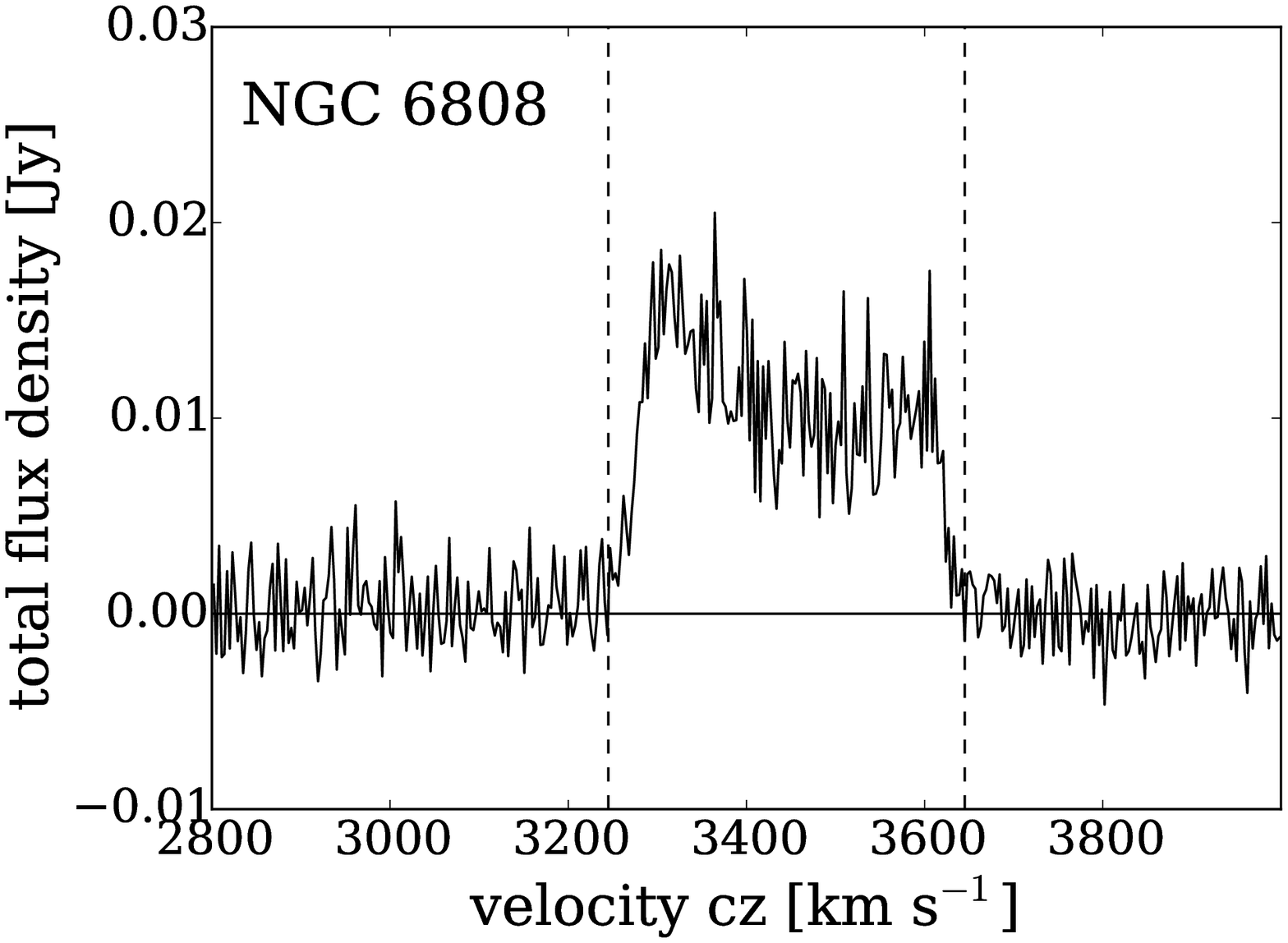}} 
	\subfigure{\includegraphics[width=5.5cm]{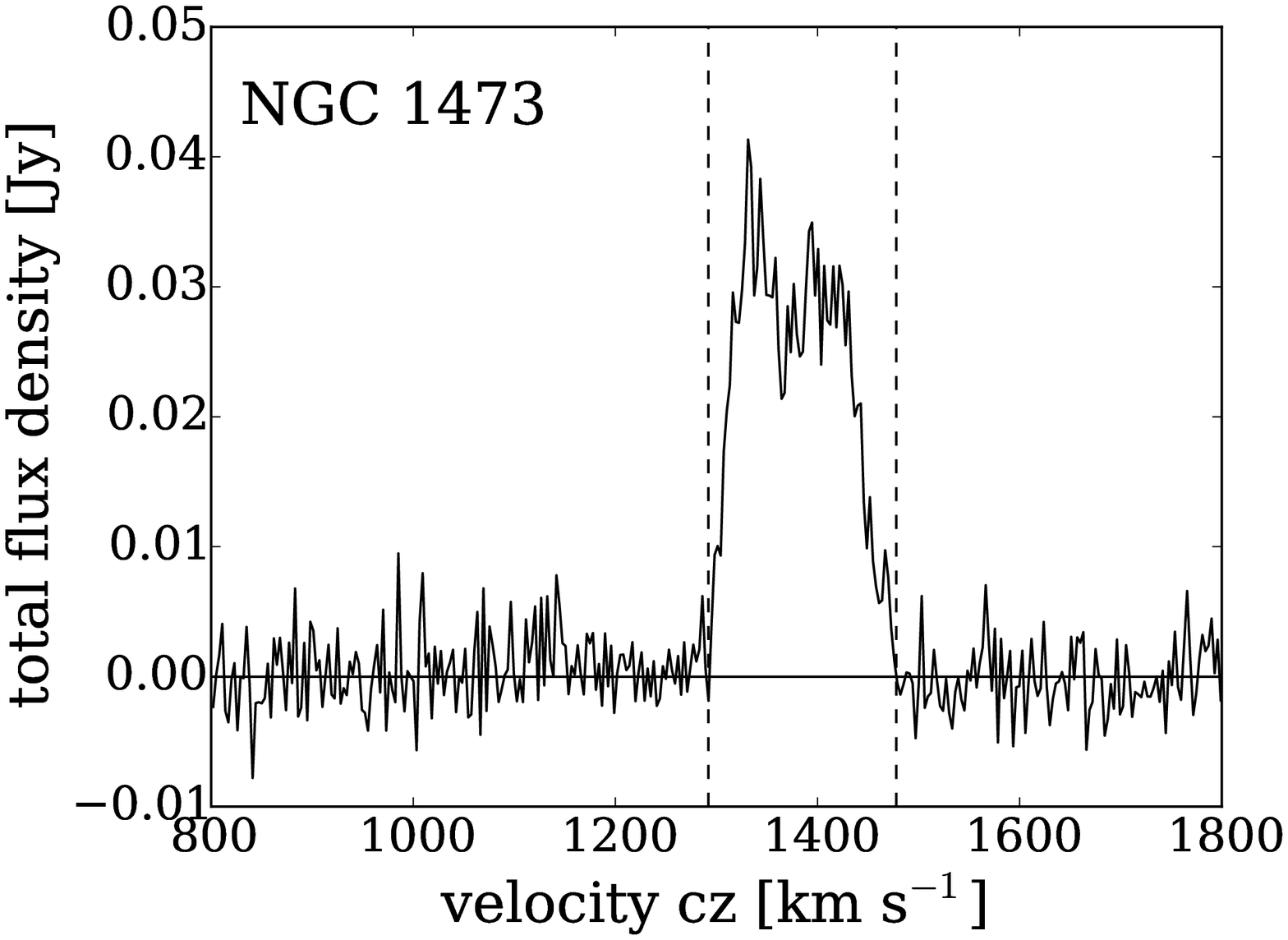}} 
	\subfigure{\includegraphics[width=5.5cm]{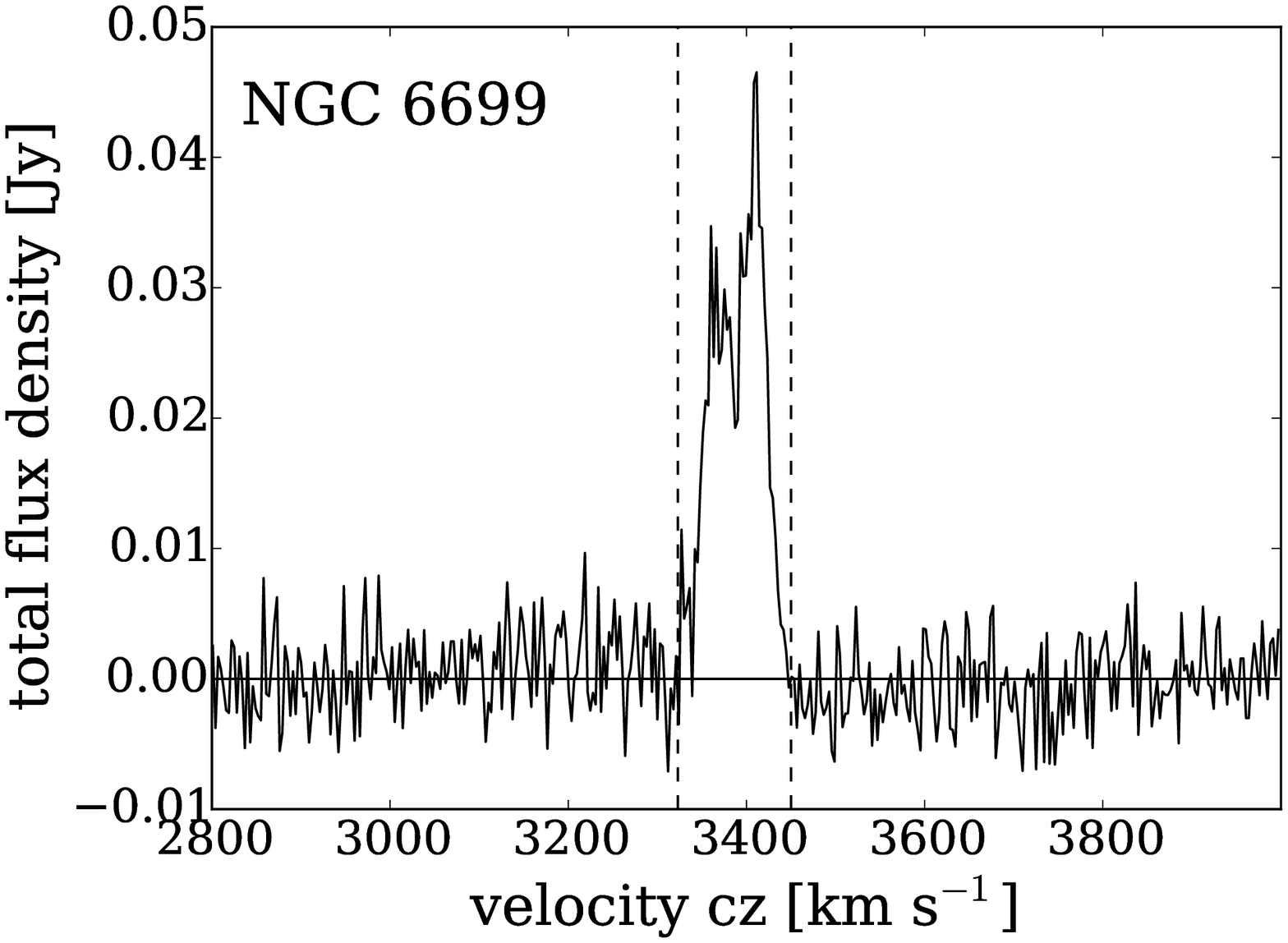}} 
	\subfigure{\includegraphics[width=5.5cm]{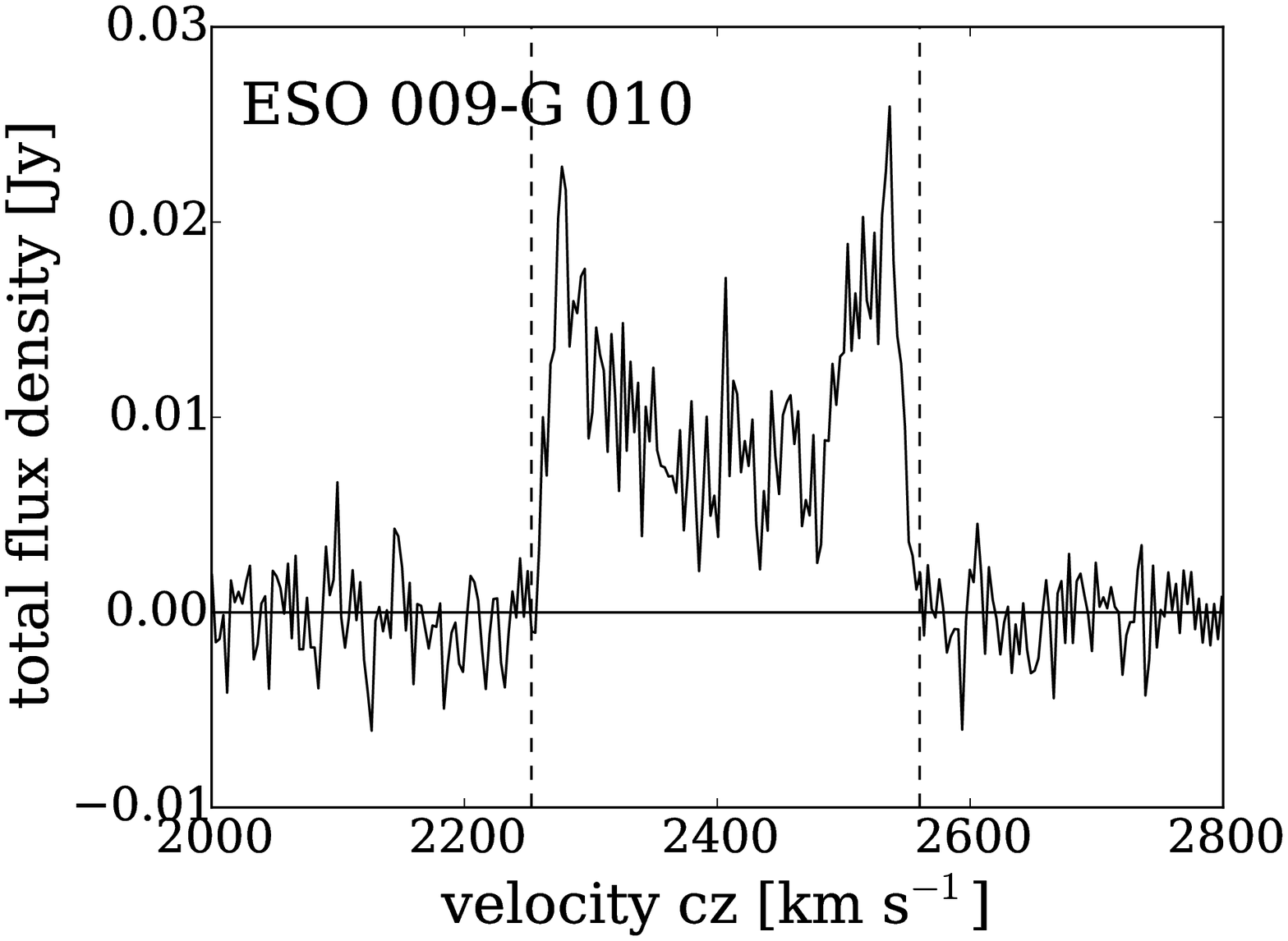}} 
	\caption{ATCA baseline subtracted \HI\ profiles. A first order baseline fit is applied to the data. Vertical dashed lines show where the \HI\ profile was integrated. }
	\label{fig:HI-profiles2}
\end{figure*}

\begin{table*}
	\caption{Summary of the gas removal signatures in each galaxy and the most likely cause of \HI-deficiency.}
	\begin{tabular}{l c c c c c c }
		\hline
		& IC 1993 & NGC 1515 & NGC 6808 & NGC 1473 & NGC 6699 & ESO 009-G 010 \\
		\hline
		Small \HI\ disk & \checkmark & \checkmark & \checkmark  & \checkmark & \checkmark & \checkmark \\
		Lopsided \HI\ disk  & \checkmark & - & \checkmark & - & \checkmark & - \\
		Asymmetric \HI\ profile  & slightly & slightly &\checkmark & \checkmark & \checkmark & slightly \\
		Diffuse \HI\ (flux ratios)  & 30 \% & - & 20\% & 25\% & 34\% & 58\% \\
		Diffuse \HI\ (\HI\ profile)  &  \checkmark & - &  \checkmark &  - &  \checkmark &  \checkmark \\
		\HI\ warp  & - &  \checkmark & - & - & - & - \\
		Kinematics & slightly pec. & - & - & - & - & slightly pec. \\
		Stellar warp  & - & - &  \checkmark & - & - & - \\
		Stellar bar & - & - & - &  \checkmark & - & - \\
		Lopsided stellar disk  & \checkmark & - & \checkmark & - & \checkmark & \checkmark \\
		Star formation rate  & avg. & avg. & starburst & avg. & increased & avg. \\
		Environment  & cluster & large group & moderate group  & small group  & moderate group & triplet \\
		Hot IGM  &  \checkmark & \checkmark &  \checkmark & - & - & - \\
		\hline  
		likely cause of    & ram pressure & ram pressure & tidal & tidal? & tidal & tidal? \\
		 \HI-deficiency: &  & & & & & \\
		\hline
	\end{tabular}
	\label{tab:signatures}
\end{table*}

\newpage

\bibliography{refs}

\end{document}